\documentclass[a4paper]{jpconf}
\usepackage{graphicx}
\usepackage[loose]{units}
\usepackage{amssymb}

\def\diffunits{\mbox{GeV cm}^{-2}\,\mbox{s}^{-1}\,\mbox{sr}^{-1}}
\def\pointunits{\mbox{GeV cm}^{-2}\,\mbox{s}^{-1}}
\newcommand{\degC}{^\circ{C}}

\begin{document}
{\Large \bf Contributions to $2^{nd}$ TeV Particle Astrophysics Conference (TeV PA II)}\\
{\Large Madison Wisconsin - 28-31 August 2006}\\
\vskip 1. cm
\begin{center}
\textbf{IceCube Collaboration}\\
\begin{paragraph}
\normalsize{
A.~Achterberg$^{31}$,
M.~Ackermann$^{33}$,
J.~Adams$^{11}$,
J.~Ahrens$^{21}$,
K.~Andeen$^{20}$,
D.~W.~Atlee$^{29}$,
J.~N.~Bahcall$^{25}$(deceased),
X.~Bai$^{23}$,
B.~Baret$^{9}$,
S.~W.~Barwick$^{16}$
% ...
}

\end {paragraph}
\vspace*{0.5cm}
\end{center}

\small{
\noindent
{$^{1}$III Physikalisches Institut, RWTH Aachen University, D-52056, Aachen, Germany}\\
{$^{2}$Dept.~of Physics and Astronomy, University of Alaska Anchorage, 3211 Providence Dr., Anchorage, AK 99508, USA}\\
{$^{3}$CTSPS, Clark-Atlanta University, Atlanta, GA 30314, USA}\\
{$^{4}$Dept.~of Physics, Southern University, Baton Rouge, LA 70813, USA}\\
{$^{5}$Dept.~of Physics, University of California, Berkeley, CA 94720, USA}\\
{$^{6}$Institut f\"ur Physik, Humboldt Universit\"at zu Berlin, D-12489 Berlin, Germany}\\
{$^{7}$Lawrence Berkeley National Laboratory, Berkeley, CA 94720, USA}\\
{$^{8}$Universit\'e Libre de Bruxelles, Science Faculty CP230, B-1050 Brussels, Belgium}\\
{$^{9}$Vrije Universiteit Brussel, Dienst ELEM, B-1050 Brussels, Belgium}\\
{$^{10}$Dept.~of Physics, Chiba University, Chiba 263-8522 Japan}\\
{$^{11}$Dept.~of Physics and Astronomy, University of Canterbury, Private Bag 4800, Christchurch, New Zealand}\\
{$^{12}$Dept.~of Physics, University of Maryland, College Park, MD 20742, USA}\\
{$^{13}$Dept.~of Physics, Universit\"at Dortmund, D-44221 Dortmund, Germany}\\
{$^{14}$Dept.~of Subatomic and Radiation Physics, University of Gent, B-9000 Gent, Belgium}\\
{$^{15}$Max-Planck-Institut f\"ur Kernphysik, D-69177 Heidelberg, Germany}\\
{$^{16}$Dept.~of Physics and Astronomy, University of California, Irvine, CA 92697, USA}\\
{$^{17}$Dept.~of Physics and Astronomy, University of Kansas, Lawrence, KS 66045, USA}\\
{$^{18}$Blackett Laboratory, Imperial College, London SW7 2BW, UK}\\
{$^{19}$Dept.~of Astronomy, University of Wisconsin, Madison, WI 53706, USA}\\
{$^{20}$Dept.~of Physics, University of Wisconsin, Madison, WI 53706, USA}\\
{$^{21}$Institute of Physics, University of Mainz, Staudinger Weg 7, D-55099 Mainz, Germany}\\
{$^{22}$University of Mons-Hainaut, 7000 Mons, Belgium}\\
{$^{23}$Bartol Research Institute, University of Delaware, Newark, DE 19716, USA}\\
{$^{24}$Dept.~of Physics, University of Oxford, 1 Keble Road, Oxford OX1 3NP, UK}\\
{$^{25}$Institute for Advanced Study, Princeton, NJ 08540, USA}\\
{$^{26}$Dept.~of Physics, University of Wisconsin, River Falls, WI 54022, USA}\\
{$^{27}$Dept.~of Physics, Stockholm University, SE-10691 Stockholm, Sweden}\\
{$^{28}$Dept.~of Astronomy and Astrophysics, Pennsylvania State University, University Park, PA 16802, USA}\\
{$^{29}$Dept.~of Physics, Pennsylvania State University, University Park, PA 16802, USA}\\
{$^{30}$Division of High Energy Physics, Uppsala University, S-75121 Uppsala, Sweden}\\
{$^{31}$Dept.~of Physics and Astronomy, Utrecht University/SRON, NL-3584 CC Utrecht, The Netherlands}\\
{$^{32}$Dept.~of Physics, University of Wuppertal, D-42119 Wuppertal, Germany}\\
{$^{33}$DESY, D-15735 Zeuthen, Germany}\\
{$^{34}$on leave of absence University of Bari, 70126, Italy}
}
\vspace*{0.1 cm}

 \pagebreak

\begin{center} 
\textbf{Table of Contents}
\end{center}
\vspace*{.5cm}
 \begin{enumerate} 

 \item Kael Hanson for the IceCube collaboration, Construction Status and Future of the IceCube Neutrino Observatory 

\item Julia K. Becker for the IceCube collaboration, Implications of AMANDA Neutrino Flux Limits

 \item D.F. Cowen for the IceCube collaboration, Tau Neutrinos in IceCube

  \item Elisa Resconi for the IceCube collaboration, IceCube: Multiwavelength Search for Neutrinos from Transient Point Sources

 \item Xinhua Bai and Thomas K. Gaisser for the IceCube collaboration, Air Showers in a Three-Dimensional Array: Recent Data from IceCube/IceTop

 \item Juan de Dios Zornoza for the IceCube collaboration, High-Energy Gammas from the Giant Flare of SGR 1806-20 of December 2004 in AMANDA

\item Jon Dumm and Hagar Landsman for the IceCube collaboration, IceCube -- First Results

  \item Jessica Hodges for the IceCube collaboration, Multi-Year Search for a Diffuse Flux of Muon Neutrinos with AMANDA-II

  \item Brennan Hughey for the IceCube collaboration, Searches for Neutrinos from Gamma-Ray Bursts with AMANDA-II and IceCube

 \end{enumerate}

\newpage
\title{Construction Status and Future of the IceCube Neutrino Observatory}
\author{Kael D. Hanson$^1$ for the IceCube Collaboration}
\address{$^1$ A3RI, University of Wisconsin - Madison, 
222 W. Washington Ave, Madison, WI 53703}

\begin{abstract} 
The IceCube neutrino telescope nears the end of its second running season having
collected a sample of over $2\times10^9$ triggered events. While the majority of
these events are cosmic ray muons, the detector is already sufficiently well
understood to allow identification of neutrino-induced muon candidate events
from the CR background. The production of optical module instrumentation is now
well-established, the modules themselves are functioning properly with low
failure rate, and it has been proven that the hot water drill can deliver the
holes needed for deployment of these instruments. The project plans to deploy
12-14 strings each year during the next several austral summers to bring the
detector volume to $\unit[1]{km^3}$.
\end{abstract}

\section{Introduction}

High-energy neutrino astrophysics is entering the era of kilometer-scale
observatories. The IceCube neutrino telescope will be the first detector with an
integrated exposure volume to reach $\unit[1]{km^3\cdot yr}$. The detector
includes a deep array of digital optical sensors deployed at depths between
\unit[1500]{m} and \unit[2450]{m} in holes drilled in the glacial ice sheet at
the geographic South Pole. These deep sensor modules detect the Cherenkov light
radiated by passing charged relativistic particles in transit through the ice
medium. The optical properties of this medium have been measured with \textit{in
situ} light sources \cite{icepaper} deployed with the predecessor detector
array, AMANDA \cite{amanda:nature, amanda:prd}: below \unit[1500]{m} the ice
becomes bubble free where long absorption and scattering lengths are found
($\ell_{\mathrm{abs}} \sim \unit[100]{m}$, $\ell_{\mathrm{scatt}} \sim
\unit[25]{m}$).

The IceCube deep array is optimized for the detection of muons produced by high
energy ($E \gg \unit[1]{TeV}$) neutrinos from astrophysical point source
emitters such as active galactic nuclei or transient sources such as gamma ray
bursts \cite{ic3:sens}. The muon is produced via charged-current interactions of
the neutrino with ice nuclei ($\nu_\mu + N \rightarrow \mu + X$), typically
exterior to the detector volume due to the long range of muons with energies in
excess of \unit[1]{TeV}. Ice is also an ideal calorimetric medium due to the
long optical absorption lengths and so the visible energy of contained neutrino
events can be reconstructed with $\pm 20$\% resolution in the exponent. In
addition to these high-energy phenomena of cosmic origin, IceCube may observe
signals from dark matter annihilations and will collect a high statistics sample
($O(10^6)$) of atmospheric neutrinos relevant to particle physics topics such as
Lorenz invariance tests in regions unreachable by other techniques. At the low
energy end, IceCube presents an effective volume of approximately
$\unit[2.0\times 10^6]{tons}$ to MeV neutrinos from supernovae.

An array of the same sensors deployed in the ice holes are frozen into tanks at
the top of each hole, providing an airshower detector component for IceCube.
Called IceTop, this instrumentation may be used as a trigger veto to 
assist in rejection of cosmic ray event backgrounds in the deep detector.
Furthermore, combining its data with data from the deep-ice array provides a 
unique opportunity to study cosmic ray composition in the region of the 
``knee,'' extending earlier measurements performed using the combination of 
the SPASE and AMANDA detectors \cite{spase-amanda, ic3:tev-icetop}.

\section{Status of IceCube instrument deployment}

The first IceCube string (\#21) and the first four IceTop stations (\#21, \#29,
\#30, and \#39) were deployed in January 2005 at the end of the deployment
season and were operated during the austral winter of that year. The
survivability of the digital optical modules during deployment and subsequent
refreeze of the drill hole was established (all DOMs deployed during this season
continue to function properly), useful performance data were gathered
throughout the year of operation of the string \cite{ic3:performance,
ic3:1st-res}, and neutrino candidate events were selected from this data run.

\begin{figure}
\begin{center}
\includegraphics[width=4.0in]{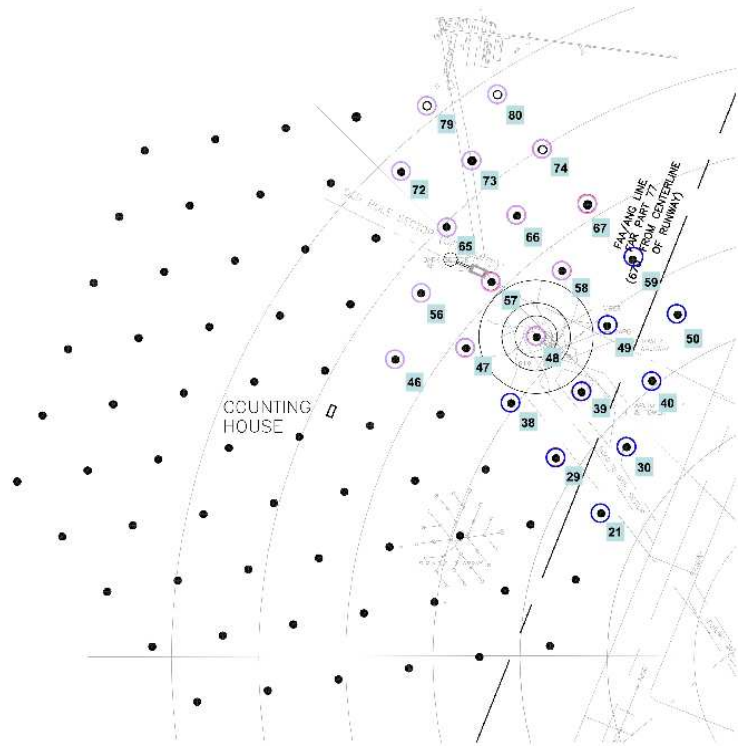}
\end{center}
\caption{Surface view showing an overlay of the IceCube detector
on the South Pole station map.  The Amundsen-Scott station is slightly
off the lower righthand corner of this illustration.  The existing 
AMANDA detector is represented by the concentric circles centered
approximately on hole \#48.  Thick dots represent planned hole locations,
those circled are either existing or planned in the 2006-2007 deployment
season.}
\label{fig:ic3-plan}
\end{figure}

During the following austral summer season, from December 2005 to January 2006,
eight more strings (\#29, \#30, \#38, \#39, \#40, \#49, \#50, and \#59) and
twelve more IceTop stations (\#38, \#40, \#47, \#48, \#49, \#50, \#57, \#58,
\#59, \#66, \#67, and \#74) were deployed bringing the count to 9 strings and 16
surface stations and a total enclosed ice volume of $\unit[0.1]{km^3}$. Of the
604 sensors deployed to date, 597 of them communicate and 592 are producing high
quality data. A current view of the IceCube detector installation is shown in
Figure~\ref{fig:ic3-plan}. The deployment plan calls for 12-14 strings and 10
surface stations to be deployed this year (2006-2007) to be followed by an
average of 14 strings and IceTop stations in the following years until 2011 when
the full complement of instrumentation will have been deployed, approximately
70-75 strings (60 DOMs per string) and 80 surface stations (4 DOMs per station).
IceCube will be operated throughout the construction, achieving an integrated
exposure of $\unit[1]{km^3 \cdot yr}$ by 2009 and $\unit[4]{km^3 \cdot yr}$ by
the second year of operation with the completed detector. We anticipate that the
total operating lifetime of the experiment will be 20 years.

\section{Drilling and deployment}
\begin{figure}[hbt]
\begin{center}
\includegraphics[width=4.0in]{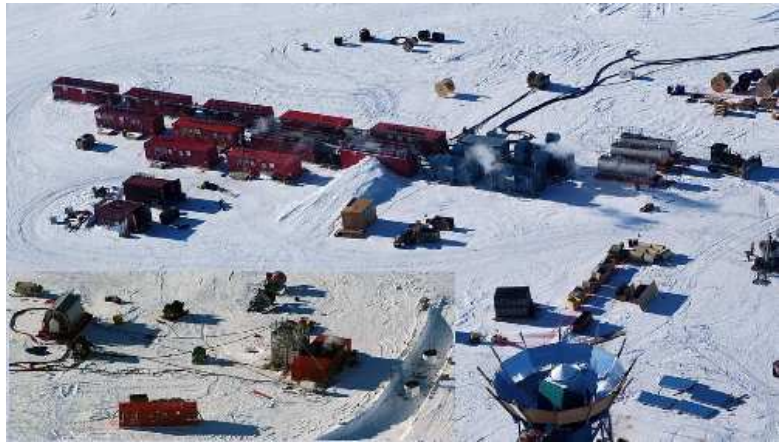}
\end{center}
\caption{IceCube drill camp with drill tower image inset in lower left corner.
The tower sits atop a drill hole.  Two icetop tanks forming a station are visible
in trench to the right of the drill tower.}
\label{fig:drillcamp}
\end{figure}

The Enhanced Hot Water Drill (EHWD) system delivers 
$\unit[2.5]{km} \times \unit[60]{cm}$ holes to the deployment team for 
insertion of the optical sensor hardware.  The system includes self-contained
heating and electrical powerplants with a combined power of approximately
\unit[5]{MW}, pumping systems, a control facility, and drilling towers.  Each
year the drill camp is moved into place near the target holes.  The towers then
operate as mobile field facilities served by the central drill camp
and towed into position atop each drill hole (Figure~\ref{fig:drillcamp}).
During operation, the drill supplies 200 gallons per minute of 
$\unit[190]{\degC}$ water at \unit[1000]{psi}.  The average fuel consumed per
hole is 7200 gallons.  The entire operation of drilling a hole and deploying
the optical module instrumentation takes approximately 50 hours.

\section{The IceCube digital optical module} 

\begin{figure}[h]
\begin{center}
\includegraphics[width=3.5in]{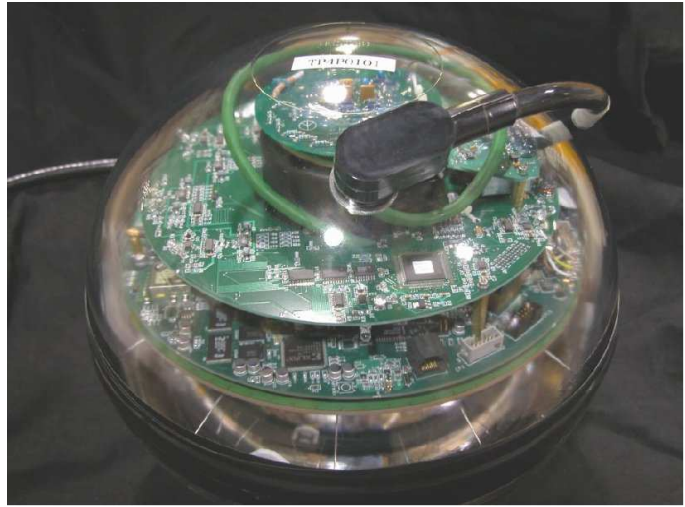}
\end{center}
\caption{
IceCube digitial optical module, shown here without mounting harness for
clarity. The PCB stack is visible with the flasher board module obscuring most
of the DOM mainboard. The HV generator module is mounted on the flasher board
(partially blocked in this figure by the penetrator assembly). The
photomultiplier tube faces downward and here is almost completely hidden under
the PCBs.
}
\label{fig:DOM}
\end{figure}

The IceCube digital optical module (DOM) (Figure~\ref{fig:DOM}) is the central
detector element used throughout the array, both in the deep ice and at the
surface. It is a self-contained optical detector and data acquisition device.
The analog optical device is a 10'' photomultiplier tube running at $1.0\times
10^7$ gain into a $\sim\unit[50]{\Omega}$ front-end load impedance. PMT high
voltage bias is supplied internally by a DC-DC converter module that is powered
from the \unit[+5]{V} line on the DOM mainboard and can produce a programmable
HV from 0 to \unit[+2048]{V}. A classical resistive divider bleeder distributes
voltages to the PMT dynodes. The DOM also contains a PCB containing 12
\unit[405]{nm} LEDs which may be flashed in the ice to provide a known optical
source for studying ice properties or performing geometrical calibrations of the
sensor array. All components are housed inside a 0.5'' thick glass pressure
sphere rated to \unit[10000]{psi} external pressure. The power and digital
communication lines exit the DOM via the penetrator cable which attaches to the
main communication cable bundles. DOM digital communication signals travel to
the surface over copper quads contained within the \unit[45]{mm} cable bundles.

DOMs are assembled at three production and test facilities worldwide within the
IceCube collaboration: University of Wisconsin, Stockholm University / Uppsala
University, and DESY Zeuthen. Following assembly each DOM undergoes a 2-3 week
test at various temperatures from $\unit[+25]{\degC}$ to $\unit[-55]{\degC}$ in
order to evaluate its performance at low temperature and to characterize various
optical and electronic operational parameters \cite{ic3:prod-test}. All data
thus far obtained with DOMs manufactured at all sites supports the claim that
all sites are producing equivalent sensor hardware. To date 2000 of a total 5000
DOMs have been built. First pass yields are nearing 90\% and the shipping yields
are in excess of 95\%.

%with a wide dynamic range from single photon hits on its 10'' photocathode 
%surface to 1000's of photons 

\section{Data acquisition} 
The PMT pulses are converted into digital waveforms
by one or more digitizer chips at speeds up to $3\times 10^8$ samples/s.
Each DOM runs in self-triggered mode with the option to monitor digital trigger
lines connected to its neighbor DOMs which it may use to influence the trigger
decision. DOM-level triggers force a digitization and readout of the digitizers
into local memory on the DOM (the DOM has a capacity of 16 MB) and each readout
is time stamped with a counter value derived from the 40 MHz local DOM
oscillator. Upon command from a surface controller, the DOM will transfer the
contents of its memory buffers to the surface at a bit rate of 1 Mbit/s per
copper pair.

At the surface, DOMs are readout by specialized PCI cards plugged into
industrial PCs running Linux.  Software running inside these computers 
must translate the DOM timestamp to a global quantity since each DOM 
oscillator is free running.  Therefore the time stamp
generated in the DOM is only locally relevant.  The time transformation is
achieved by a process called RAPCal wherein the DOM and the surface
digital communication hardware periodically (approximately once per second)
exchange analog pulses and stamp the arrival and departure times.  This 
information is used to establish the DOM clock to surface clock mapping.  The
clocks at the surface are driven from a single 10 MHz master clock signal
synchronized to GPS.  Measurements in the laboratory and \textit{in situ} at
South Pole demonstrate that DOM-to-DOM time jitter is $O(\unit[3]{ns})$
less than the design specification of \unit[5]{ns}.

Once the digitized PMT pulses have been stamped with a global time, they are
merged and sorted into a stream which is sent over ethernet to a cluster of
trigger and event processor computers. The triggering and event packaging is
accomplished entirely in application software. During the 2006 run, two triggers
were implemented: a minimum bias trigger (MBT) generating an event
trigger every $n$-th hit for system debugging and the main trigger 
for physics analysis, the simple majority trigger (SMT), requiring
coincidence of 8 or more DOMs hit in the deep-ice array or 6 or more hits in the
IceTop array within a time window of $\unit[5]{\mu{s}}$. The triggers were
formed in separate trigger processors for the in-ice and IceTop arrays;
coincident triggers were then handled by a global trigger unit. Typical trigger rates
from a run in mid-winter operation are listed in Table~\ref{tab:trigger-rates}.

\begin{table}[hbt]
\begin{center}
\begin{tabular}{|c|c|c|} \hline
 & MBT & SMT \\ \hline
In-Ice & \unit[5.28]{Hz}  & \unit[139]{Hz} \\ \hline
IceTop & \unit[0.875]{Hz} & \unit[6.43]{Hz} \\ \hline
\multicolumn{2}{|c|}{IceTop - In-Ice Coincident} & \unit[0.25]{Hz} \\ \hline
\end{tabular}
\end{center}
\caption{Trigger rates from a June 23, 2006 data run.}
\label{tab:trigger-rates}
\end{table}

\section{Summary}
IceCube is soon to begin its $3^{\mathit{rd}}$ deployment season after
concluding a successful deployment and running season.  All indications
from data quality verification studies point to the hardware functioning
at or above its design specification.  The detector will reach an
integrated exposure volume of $\unit[1]{km^3 \cdot yr}$ in as little as
two years' time.  Future detectors involving acoustic and radio detection
techniques are being investigated as potential additions to the IceCube
observatory to substantially extend the detector volume, particularly 
at higher energies.

\newpage
\setcounter{section}{0}

\title{Implications of AMANDA neutrino flux limits}

\author{Julia Becker for the IceCube
  Collaboration\footnote{http://icecube.wisc.edu}}

\address{Universit\"at Dortmund, Institut f\"ur Physik, 44221 Dortmund, Germany}

\ead{julia.becker@udo.edu}

\begin{abstract}
The Antarctic Muon And Neutrino Detector Array (AMANDA) is currently the
most sensitive neutrino telescope at high energies. Data
have been collected in a period of eight years and analyzed with different analysis strategies.
Limits to the neutrino flux from point sources, transient emissions, source catalogs
and limits to different diffuse flux models have been obtained implying in
some cases strong contraints to hadronic interaction models of such sources. In this
contribution, implications of the diffuse neutrino limit will be discussed with respect to
neutrino production mechanisms in astrophysical sources.
\end{abstract}
%%%%%%%%%%%%%%%%%%%%%%%%%%%%%%%%%%%%%%%%%%%%%%
\section{Neutrino flux predictions}
\begin{figure}[h]
\begin{minipage}{18pc}
\includegraphics[width=20pc]{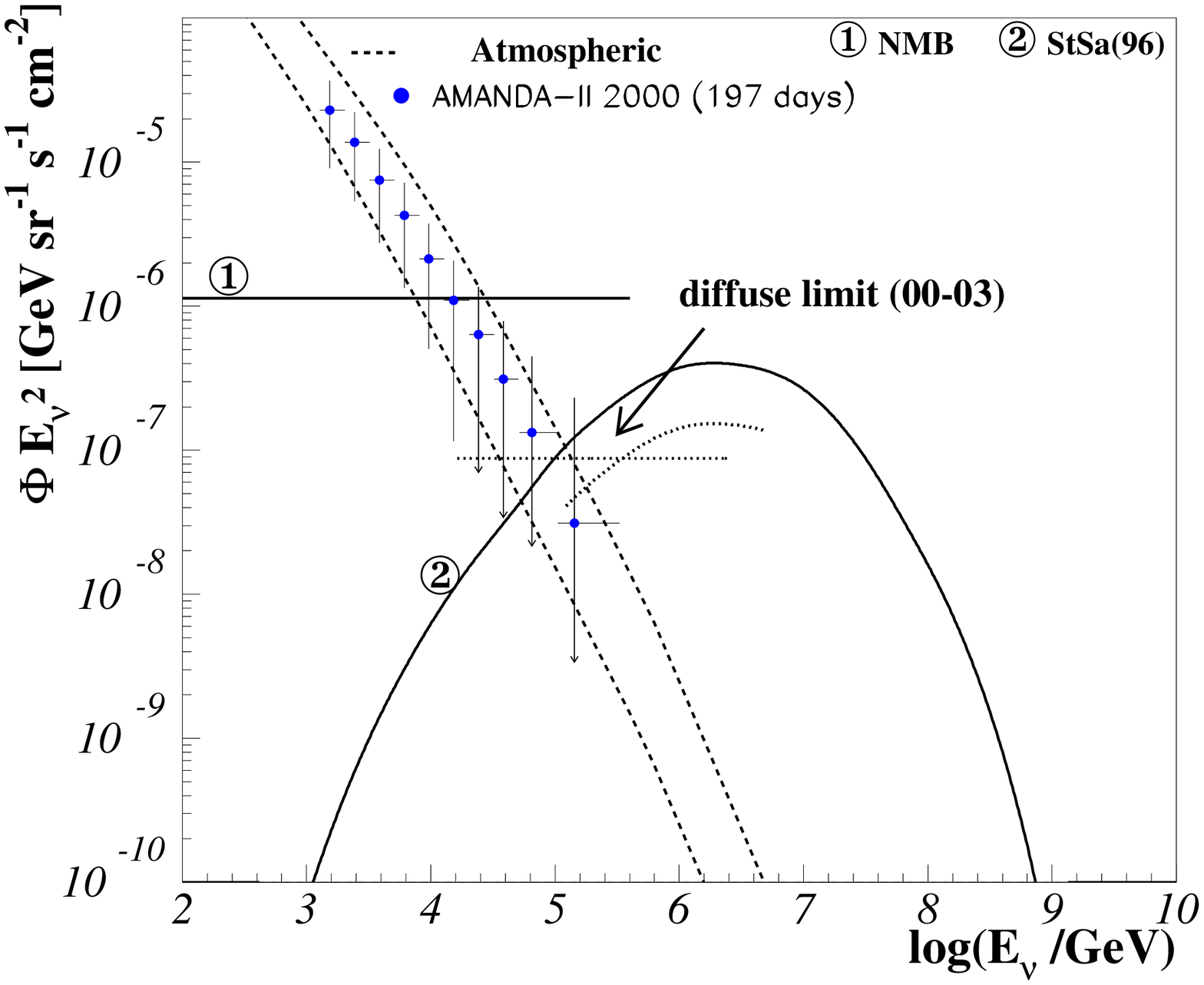}
\end{minipage}\hspace{2pc}%
\begin{minipage}{18pc}
\includegraphics[width=20pc]{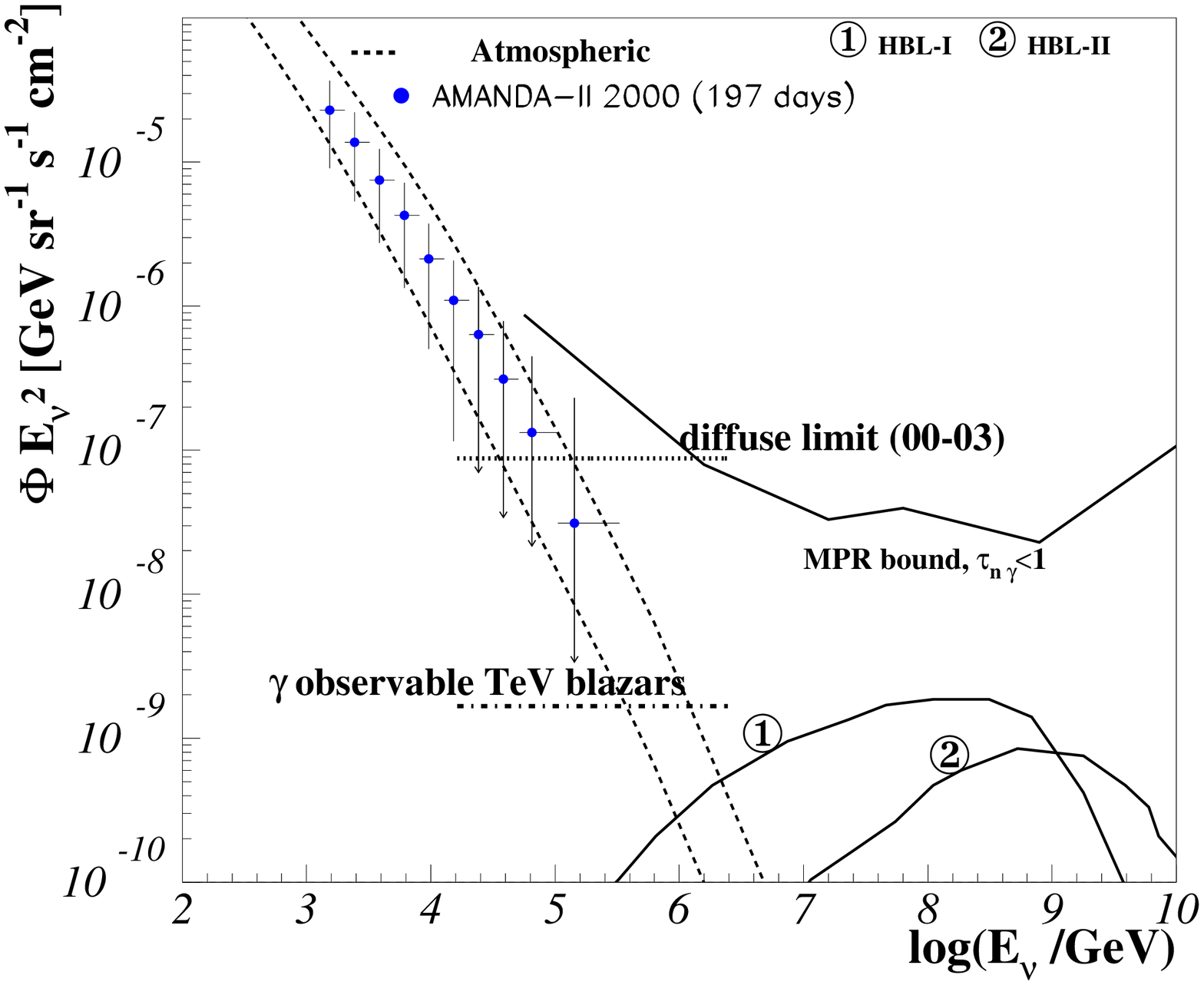}
\end{minipage} 
\caption{\label{spectra}$\nu$ spectra for models of $\nu$ emission from
    X-ray emitting AGN (left panel) and for optically thin sources -
    the cosmic ray flux at the highest energies is assumed to be proportional
    to the $\nu$ output (right panel). Data points are measurements of the diffuse spectrum
    by AMANDA, year 2000~\cite{jess_these_proc}. Dashed lines represent the
    atmospheric contribution, the lower line is the vertical flux, the upper
    one represents the horizontal flux. Limits for 4 years
    ('00-'03, lifetime=807 days), dotted lines.  AGN $\nu$ predictions: (1)
    from Ref.~\cite{nellen}; (2) from~\cite{stecker96} (left panel) and (1)
    and (2) from Ref.~\cite{muecke} (right
    panel). On the right, the maximum contribution of $\gamma$ observable blazars is
    shown as the dot-dashed line. Flux predictions account for $\nu$
    oscillations.}
\end{figure}

%%%%%%%%%%%%%%%%%%%%%%%%%%%%%%%%%%%%%%%%%%%%%%
The existence of Ultra High Energy Cosmic Rays (UHECRs) as well as the
detection of TeV photon emissions from galactic and extragalactic sources are a
strong indication for neutrino ($\nu$) emission from the same sources. Pions and kaons
are believed to take a fraction of the proton energy producing TeV photons in
coincidence with high energy neutrinos. Although the atmospheric background of neutrinos is quite high,
it decreases rapidly with energy ($\sim E^{-3.7}$) while the extraterrestrial
spectra of galactic and extra-galactic sources are typically flatter (typically $\sim E^{-2}$ if shock acceleration
is the main mechanism producing high energetic protons at the source). The latter should therefore become the
dominant component of the total diffuse spectrum at a certain energy, which
depends on the normalization of the neutrino flux. Different predictions are
shown in Fig.~\ref{spectra}. The left panel shows various calculations which
use the diffuse X-ray background as measured by ROSAT to normalize the
neutrino spectrum, see~\cite{stecker96,nellen}. This is justified when assuming the production of
neutrinos along with X-rays at the foot of jets of Active Galactic Nuclei (AGN) where protons are
accelerated into the photon target of the disk. The right panel shows models
based on the correlation between UHECRs, TeV photons and neutrinos, see~\cite{mpr,muecke}. Such
sources are optically thin to both TeV photons and protons.
%%%%%%%%%%%%%%%%%%%%%%%%%%%%%%%%%%%%%%%%%%%%%%
\section{Detection techniques of AMANDA}
%%%%%%%%%%%%%%%%%%%%%%%%%%%%%%%%%%%%%%%%%%%%%%
AMANDA detects muon-neutrinos ($\nu_{\mu}$s) by observing secondary muons
from charged current interactions of the neutrinos with the nucleons of the
ice. The muons are traveling faster than light in ice and emit Cherenkov
radiation which is detected by the photomultiplier tubes. Between the
years 2000 and 2004 data from effectively 1001 days have been taken and a $\nu_{\mu}$ sample
of 4282 events from the Northern hemisphere has been
collected\footnote{Atmospheric muons make it impossible to use the Southern
  hemisphere for $\nu_{\mu}$ searches. Cascade analyses can, however, be done
  for both hemispheres. These results will not be discussed here, but can be
  found in e.g.~\cite{icrc05}}. 
In order to keep the analysis blinded to avoid experimenter's bias, analyses
cuts are optimized using off-source samples created by scrambling the right
ascension of events or excluding the time window of transient emissions under
investigation. For the case of diffuse flux analyses the analysis is optimized
on a low energy sample, where the signal is expected to be negligible.

AMANDA has a twofold stragegy for searching for steady point sources. In a
first  method, a source catalog of 32 sources was established and spatial
cuts were determined based on the position of the potential neutrino emitters. The second technique searches for the spatial clustering of
events. Neither of the two point source searches has shown a significant
excess of events. 
The mean sensitivity in the Northern hemisphere to an $E^{-2}$ neutrino flux is
\begin{equation}
E^2\Phi_{lim}=5.9\cdot 10^{-8}\pointunits
\end{equation}
for 5 years of data taking.
Here, E is the neutrino energy and $\Phi_{lim}$ is the flux upper limit.

The search for single point sources was complemented by stacking classes of
sources according to the direct correlation between the photon output and the
potential neutrino signal. This was done for 11 different AGN samples
that were selected at different wavelength bands, see~\cite{andreas_stacking}. The
optimum sensitivity was typically achieved by the stacking of around 10
sources. The cumulative and mean source limit for every class is given in table~\ref{stacking_table}.

\begin{table}[!h]
\caption{\label{stacking_table}Results of the stacking analysis for
each AGN category~\cite{point_source_5yr}:
the number of included sources $N_{src}$, the number of
   expected events $N^{bg}_\nu$ and  the number of observed
   events $N^{obs}_\nu$ are listed as well as the cumulative limit $f_{lim}$
   as well as the limit per source $f_{lim}/N_{source}$, both in units of $\pointunits$.
}
\begin{center}
\begin{tabular}{l|llllll}
\br
AGN category & $N_{src}$ & $N_{\nu}^{obs}$  & $N_{\nu}^{bg}$
&$f_{lim}$& $f_{lim}/N_{src}$ \\
\mr
GeV blazars & 8 & 17 & 25.7&2.71&0.34 \\
unidentified GeV sources& 22 & 75  & 77.5&31.7&0.75\\
IR blazars & 11 & 40 & 43.0 &10.6&0.96\\
keV blazars (HEAO-A) & 3 & 9 & 14.0&3.55&1.18\\
keV blazars (ROSAT)  & 8 & 31 & 33.4&9.71&1.2\\
TeV blazars & 5 & 19 & 23.6&5.53&1.11\\
GPS and CSS & 8 & 24 & 29.5&5.94&0.74\\
FR-I galaxies & 1 & 3 & 3.1&4.11&4.11\\
FR-I without M87 & 17 & 40 & 57.2&2.91&0.17\\
FR-II galaxies& 17 & 77 & 68.5&30.4&1.79\\
radio-weak quasars& 11 & 35 & 41.6&6.70&0.61\\
\br
\end{tabular}
\end{center}
\end{table}

In the diffuse analysis high-energy (HE) events from all directions are examined
with respect to the spectral energy behavior of the sample. A flattening of
the total neutrino spectrum is expected when a flat, astrophysical component
($\Phi\sim E^{-2}$) overcomes the steep atmospheric background ($\Phi\sim
E^{-3.7}$). The reconstructed energy spectrum for one year of data (year 2000)
is shown in Fig.~\ref{spectra}. It follows the atmospheric prediction (dashed
lines). The most restrictive limit from the diffuse analysis for the years
2000 to 2003 is given as
\begin{equation}
E^2\,\Phi_{lim}=8.8\cdot 10^{-8}\diffunits
\end{equation}
in the energy range of $4.2<\log(E/\mbox{GeV})<6.4$.

The results were obtained by optimizing the analysis cuts on
$E^{-2}$ spectra. Nonetheless the dependency of the response function of the
detector to different spectra was considered and limits were set for different
spectral shapes (e.g.~$E^{-3}$) or specific models as shown in Fig.~\ref{spectra}.
Varying the spectral index in
the simulation shows that the event distribution simulated for AMANDA peaks at
very different energies depending on the assumed spectral index. While, for an
$E^{-2}$ spectrum, $90\%$ of the signal lies between
$4.2<\log(E/\mbox{GeV})<6.4$ as discussed above, an $E^{-3}$ spectrum shows an
event distribution located about an order of magnitude lower in energy while
an $E^{-1}$ spectrum shifts the sensitivity to higher energies. This shows
that it is useful to model the spectra according to the predicted shape. This
is discussed in detail in~\cite{jess_these_proc}.
%%%%%%%%%%%%%%%%%%%%%%%%%%%%%%%%%%%%%%%%%%%%%%
\section{Interpretation of AMANDA diffuse limit}
%%%%%%%%%%%%%%%%%%%%%%%%%%%%%%%%%%%%%%%%%%%%%%
Two main astrophysical implications to be drawn from the current diffuse limit
will be examined here. The first is the apparent overproduction of neutrinos
in coincidence with X-ray photons in the case of hadronic acceleration at the
foot of AGN jets. The second is the maximum contribution of TeV observable
blazars to the total diffuse neutrino flux. For a detailed discussion of these
and further implications, see~\cite{diffuse_part}.
\subsection{X-ray/Neutrino correlation in AGN}
The left panel of Fig.~\ref{spectra} shows that two models, predicting
neutrino emission from X-ray emitting AGN, violate the AMANDA limit. In the
case of model~1~\cite{nellen}, the $E^{-2}$-shaped limit applies (constant, dotted curve), while the
limit has been calculated according to the specific shape of the model in the
case of model~2~\cite{stecker96} (curved, dotted line). 
Another model~\cite{alvarez_meszaros} relating neutrino to X-ray emission 
can be ruled out in the same way. This suggests that the
observed X-rays are related to Inverse Compton Scattering
rather than to a hadronic scenario.
This, however, does not rule out neutrino emission in coincidence
with other wavelength bands, like MeV, GeV or TeV sources, for example.
\subsection{TeV blazars and Neutrinos}
The diffuse neutrino flux from TeV blazars must be lower than the 
diffuse AMANDA limit:
\begin{equation}
E^2\left.{\frac{dN}{dE}}\right|_{TeV}<8.8\cdot 10^{-8}\diffunits\,.
\label{all_dl}
\end{equation}
Since TeV photons are absorbed on their way to Earth, current TeV Air
Cherenkov telescopes can only detect sources up to $z<0.3$. TeV photons are
believed to be directly correlated to HE neutrinos, since both are
produced via the pions from the $\Delta$-resonance resulting from $p\,\gamma$
interactions. Thus, the detected neutrino flux from {\it TeV observable
  sources} is
\begin{equation}
E^2\left.{\frac{dN}{dE}}\right|_{TeV obs}<8.8\cdot 10^{-8}\cdot\eta^{-1}\diffunits\,.
\label{abs_dl}
\end{equation}
Here, $\eta$ is the absorption factor, depending on the Star Formation Rate (SFR)
scenario and on the maximum redshift, i.e.~$z_{\max}=0.3$. Using a constant density of sources, $\eta$ is
maximized and a limit of $\eta(z_{\max}=0.3)>53$ is given. Thus, the upper
limit of the contribution of TeV observable sources to a diffuse neutrino flux
is given as 
\begin{equation}
E^2\left.{\frac{dN}{dE}}\right|_{TeV obs}<1.7\cdot 10^{-9}\diffunits\,,
\end{equation}
displayed as the dot-dashed line in the right panel of Fig.~\ref{spectra}.
This underlines the necessity of a diffuse search with HE neutrino
telescopes and the need for source-catalog independent searches. 
%%%%%%%%%%%%%%%%%%%%%%%%%%%%%%%%%
\section{Conclusions}
%%%%%%%%%%%%%%%%%%%%%%%%%%%%%%%%%
Currently, AMANDA is the most sensitive neutrino telescope at high
energies. Limits from 5 years for the point source analysis and four years for
the diffuse analysis can already be used to constrain the
physics of X-ray emission in AGN. Other acceleration mechanisms, predicting
the emission of neutrinos in coincidence with TeV, GeV or MeV photons, are
still very interesting to look for and represent interesting targets for
observation. Optically thin sources are only observable in TeV photons up to
$z<0.3$, which leaves neutrinos as a unique messenger from higher
redshifts. IceCube is currently being built at the South Pole as AMANDA's $1$km$^3$-successor and the sensitivity will reach levels of
$E^2\,\Phi_{sens}\sim(2-7)\cdot 10^{-9}\diffunits$ in only one year of full
observation, see e.g.~\cite{kael_these_proc,francis_barca}. 
This will allow to constrain further neutrino emissions from extragalactic
sources, such as AGN and Gamma Ray Bursts (GRBs), or galactic sources such as micro-quasars and
Supernova Remnants.
\ack{
Acknowledgments from the IceCube collaboration can be found at \verb+http://icecube.wisc.edu+.
The author thanks the BMBF for the possibility to attend the conference (grant: 05 CI5PE1/0).}

\newpage
\setcounter{section}{0}
\title{Tau Neutrinos in IceCube}

\author{D.F. Cowen for the IceCube Collaboration}

\address{Physics Department, 104 Davey Laboratory, Pennsylvania State University, University Park, PA 16802  USA}

\begin{abstract}
Tau neutrino detection in IceCube would be strong evidence for the presence of cosmologically-produced neutrinos.  In addition to the well-known ``double bang'' signature, we describe here five additional channels that we believe will not only extend the energy range over which IceCube can be sensitive to tau neutrinos, but also provide useful control over systematic uncertainties via self-consistency checks amongst all detection channels.
\end{abstract}

\section{Introduction}
In the search for ultrahigh energy neutrinos of cosmological origin, few pieces of evidence would be more convincing than a cleanly identified high energy tau neutrino.  Tau neutrinos are not produced in standard cosmic-ray atmospheric interactions that create electron and muon neutrinos, and they are expected at immeasurably small levels in the prompt neutrino flux created in charm particle decays in cosmic-ray interactions at high energies \cite{thun1996}.  Furthermore, at the energy and distance scales relevant for IceCube detection of atmospheric neutrinos, oscillations of $\nu_{e}$ and $\nu_{\mu}$ into $\nu_{\tau}$ will be very limited and will not result in large numbers of $\nu_{\tau}$'s at the detector.  After ruling out all these possible high energy $\nu_{\tau}$ sources, the only one left is a cosmological source that produces $\nu_{e}$ and $\nu_{\mu}$ that oscillate over large travel distances to produce a measurable number of $\nu_{\tau}$'s at the detector.

The standard UHE neutrino production mechanism is charged pion (and kaon) decay.  Pion decay makes a neutrino beam with a flavor ratio of $\nu_{e}:\nu_{\mu}:\nu_{\tau}=1:2:0$.  It is expected that neutrino oscillations will result in a 1:1:1 flavor ratio at the detector, and large deviations from this ratio would be an indication of new or unexpected physics, either in the production mechanism at the source, the propagation of the neutrinos over cosmological distances, or in the neutrino oscillation mechanism itself \cite{kash2006}.  Likewise, excessive $\nu_{\tau}$'s at atmospheric neutrino energies might also be an indicator of new physics, but IceCube will have limited ability to exclusively identify $\nu_{\tau}$'s at these lower energy scales, where a cascade from a $\nu_{\tau}$ will be very difficult to distinguish from a cascade from a charged-current $\nu_{e}$, a neutral-current any-flavor neutrino, or a low energy charged-current $\nu_{\mu}$ interaction.

\section{Tau Neutrino Signatures in IceCube}
By virtue of the tau lepton's long decay length at ultrahigh energies, and its wide variety of decay modes, a tau produced in a charged-current $\nu_{\tau}$ interaction has a rich set of possible signatures in the IceCube detector.  The tau decay length is about 50 m per PeV, so a tau with $E_{\tau}$ up to about 20 PeV can be fully contained in the detector volume.  More generally, the tau production vertex, decay vertex or both may be observable in a single event.  The tau can decay leptonically, $\tau  \rightarrow e\nu_{e}\nu_{\tau}$ (branching ratio = $\sim$18\%) or $\tau \rightarrow \mu\nu_{\mu}\nu_{\tau}$ ($\sim$18\%), or hadronically, mainly to charged and neutral pions and kaons ($\sim$64\%).  Note that since the average charged-current $\nu_{\tau}$ interaction produces a tau with 0.75$E_{\nu}$, we will assume $E_{\nu} = E_{\tau}$ and refer to either as just ``E'' for the sake of simplicity.

The following subsections and Figure 1 list six tau neutrino signatures to which IceCube may be sensitive.  For each signature we also describe the chief expected backgrounds, energy range over which IceCube will have sensitivity, relevant tau branching ratio, rough IceCube angular acceptance, and energy and pointing resolution relative to that expected for $\nu_{e}$ and $\nu_{\mu}$ events.  In order to assure that we can distinguish a tau track from one or both of its cascades, or from a muon track, we require a tau track length of at least 200 m in the detector.  However, energetic downward-going taus will encounter a higher number of DOMs (Digital Optical Modules) \cite{astro2006} per unit track length, decreasing the minimum required track length and hence the energy threshold.  Likewise, some taus will simply live longer, also lowering the energy threshold somewhat.  Detailed Monte Carlo studies of IceCube sensitivity versus track length will change our simplistically sharp 200 m cutoff.

\subsection{Double Bang}
The classic $\nu_{\tau}$ signature is the ``double bang'' in which the initial charged-current interaction creates a hadronic shower and a tau lepton \cite{lear1995}.  The tau lepton has sufficient energy to travel a long enough distance in the detector such that when it decays ($\tau \rightarrow e\nu_{e}\nu_{\tau}$  or  $\tau \rightarrow$ hadrons $\nu_{\tau}$, total BR = $\sim$82\%), it produces a second, separately visible shower.  The tau lepton connecting the two showers will also emit Cherenkov light.  Depending on the length of the tau track, and the extent to which the two showers are contained, IceCube can get the best of both worlds when reconstructing such an event: the pointing resolution can be comparable to that of $\nu_{\mu}$, and the energy resolution to that of $\nu_{e}$.  IceCube should have slightly more than 2$\pi$ sr acceptance in this channel, since upward-going $\nu_{\tau}$'s are either absorbed or degraded in energy by passage through the earth.  There should be negligible background, although in principle downward-going signal events could be faked by coincident muons from cosmic-ray air showers.  The fake rate is probably too small to be a concern, but Monte Carlo studies are needed to verify this.  The requirements that the two showers are well-separated (more than $\sim$100 m apart) and contained give a $\nu_{\tau}$ energy acceptance range of E $\sim$ 2-20 PeV.

\subsection{Lollipop}
If the tau lepton is created sufficiently far from the fiducial volume that the initial hadronic shower is not visible by the detector, and the tau then enters the fiducial volume and decays to produce a shower (total BR = $\sim$82\%), the event signature resembles a lollipop: a track ending in a shower.  The pointing resolution of lollipop events should be comparable to that of $\nu_{\mu}$, while the energy resolution will be better but not as good as for $\nu_{e}$ on account of the missed initial shower.  Requiring that the tau have at least 200 m of length in the fiducial volume gives an energy acceptance range of $E  \gtrsim  5$ PeV.  At this energy scale we are restricted to $\sim$2$\pi$ sr acceptance since upward-going neutrinos are either absorbed or degraded in energy by passage through the earth.  This channel may be sensitive to background from coincident muons from cosmic-ray air showers.  Again, this background rate would need to be estimated from Monte Carlo.

\subsection{Inverted Lollipop}
If the tau lepton is instead created inside the fiducial volume and then decays undetectably outside the volume, the signature also resembles a lollipop, but created in inverse order (shower first, track second).  While there is no branching ratio factor here--100\% of the $\nu_{\tau}$ that experience a charged-current interaction in the detector volume will produce a tau in the detector volume--and while the pointing and energy resolutions are comparable to that of the non-inverted lollipop, the inverted lollipop is susceptible to an irreducible background from $\nu_{\mu}$ charged-current events.  (A $\nu_{\mu}$ can have a charged-current interaction in the fiducial volume, creating a shower and a muon, the combination of which will be very hard to distinguish on an event-by-event basis from $\nu_{\tau}$  inverted lollipops.  Of course, the event itself is still of interest, especially if its energy is high enough to make it a candidate for being of cosmological origin, but here we are concerned mainly with events that we can convincingly identify as coming from a $\nu_{\tau}$.)  This channel is also susceptible to background from cosmic-ray muons accompanied by a bremsstrahlung interaction, a background that may be studied (but not eliminated) using an IceTop-tagged muon beam in the data.  As with the lollipop signature, requiring that the tau have at least 200 m of length in the fiducial volume gives a energy acceptance range of $E \gtrsim 5$ PeV.  As in the double bang and lollipop channels above, this energy scale also restricts this channel to $\sim$2$\pi$ sr acceptance.

\subsection{Sugardaddy}
If a tau is created well outside the fiducial volume, then enters it and decays to a muon rather than a shower inside the volume (BR = $\sim$18\%), it creates a unique signature which may be detectable.  As described in Ref.~\cite{deyo2006}, the much heavier tau will emit significantly less light along its length compared to its lighter daughter muon.  This can be idealized as a step-function change in track brightness.  For energies between roughly 1 PeV and 1 EeV, the magnitude of the change is expected to be larger than the track energy resolution of IceCube, and hence it should be detectable.  There is no background aside from the very unlikely high energy muon that by random chance has significantly more stochastic light-producing interactions at later times relative to earlier times along its length.\footnote{This potential background may be conservatively estimated from data by measuring the probability for tracks to have a step-function-like change in brightness but in the opposite sense from that expected for tau decays, i.e., brighter then dimmer.}   The energy resolution should be comparable to that of a standard $\nu_{\mu}$ event.  Requiring at least 200 m of tau track in the fiducial volume sets the energy scale for these events at $E >5$ PeV.

\subsection{Double Pulse}
At energies below the scale at which the two showers of a double bang can be resolved as two separate cascades by the full detector, there is an energy range in which one or more DOMs near the two closely-spaced showers will see a double-peaked structure in its waveform.  The lower end of the energy acceptance range for this signature is defined by the ability of a DOM to resolve two waveforms.  We assume here that the two waveforms need to be separated in time by $\sim$20 ns, from which we infer a lower energy of $\sim$100 TeV, corresponding to roughly a 5 m tau decay length.  Assuming the popular $E^{-2}$ spectral shape of the neutrino signal flux, this order-of-magnitude improvement in energy sensitivity relative to the standard double bang signature may be a great benefit, even after taking into account the phase space factor reduction in acceptance due to the need for favorable spacing and orientation of the two showers relative to the DOM(s).  Note that there will be slivers of phase space where the tau direction is highly favorable for creating resolvable waveforms, possibly decreasing the lower end of the energy acceptance range by an additional factor of 2 to 3.  Monte Carlo studies are underway to map out this phase space in energy and tau orientation.  The chief background to this signal would come from two successive bremsstrahlung interactions in a downward-going cosmic-ray muon event.  Presumably this background can be addressed simply by removing events with track-like topologies and requiring that all events are well contained within the fiducial volume.

\subsection{Tautsie Pop}
When a tau is at such low energy that its production and decay vertices are indistinguishable in IceCube, the decay $\tau \rightarrow  \mu \nu_{\mu} \nu_{\tau}$ will produce an inverted-lollipop--like signature in the detector and may give us access to very low tau neutrino energies.  The additional two neutrinos in the tau decay are the key difference between this signal and its backgrounds from, for example, charged-current $\nu_{\mu}$ interactions.  These neutrinos will carry away energy, causing the ratio $E_{shower} / E_{track}$ to be larger by a factor of 2 to 3 in tau events than in background events.  Due to event-to-event variations in $E_{shower} / E_{track}$, and due to the relatively small ratio difference factor, this analysis would have to be done on a statistical basis.  Although this channel suffers from the 18\% $\tau \rightarrow \mu\nu_{\mu}\nu_{\tau}$ branching ratio, it benefits from its reach to very low energies, starting at roughly where the ``double pulse'' topology leaves off and extending down to energies perhaps as low as tens of TeV.  The lower bound is determined by the energy at which the emerging muon track travels too short a distance to get a good handle on its energy, although ideally one could get to very low muon energies using events in which the muon decays in the detector volume.  At low enough energies, tau neutrinos from atmospheric interactions may start entering as a ``background'' to cosmological tau neutrinos.  In principle, background from cosmic-ray muons that have a fortuitous bremsstrahlung interaction could be estimated from a sample of IceTop-tagged muons in the data.  Background from charged-current $\nu_{\mu}$  interactions would have to be estimated from Monte Carlo.

\section{Conclusion}
\begin{figure}[htb]
\begin{center}
\includegraphics[width=3.5in]{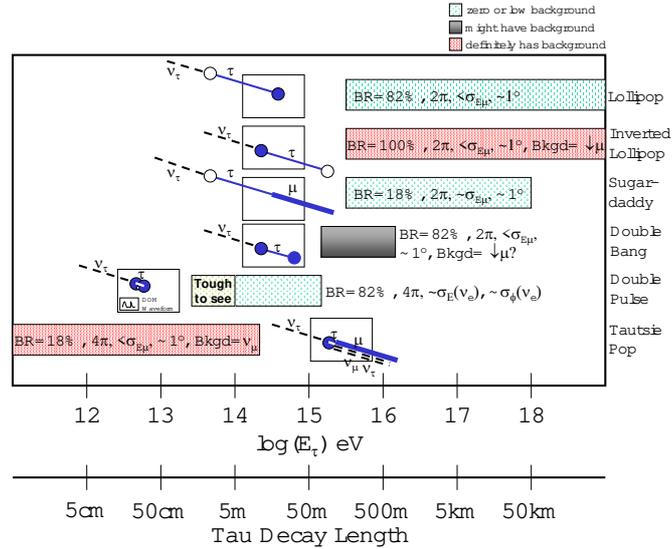}
\end{center}
\caption{\label{summ}
Summary of $\tau$ channels possibly accessible to IceCube, shown as a function of energy and approximate tau decay length, with indications of background level, acceptance, angular and energy resolutions, and specific anticipated background.}
\end{figure}

Figure \ref{summ} summarizes the tau decay channels that may be accessible to IceCube.  In addition to the canonical double bang channel, five other channels may also be detectable.  The energy range is extended considerably beyond that available from just the double bang channel alone, and in many cases the overlapping energy ranges will permit IceCube to make simultaneous tau neutrino flux measurements using channels with very different systematics.

\newpage
\setcounter{section}{0}
\title{IceCube: Multiwavelength Search for Neutrinos from Transient Point Sources}

\author{Elisa Resconi for the IceCube Collaboration}

\address{Max-Planck-Institute for Nuclear Physics, Saupfercheckweg 1, 69117 Heidelberg, Germany.}

\ead{elisa.resconi@mpi-hd.mpg.de}

\begin{abstract}
In this paper we discuss the strategy developed in order to associate neutrinos with their cosmic sources
using historical light curves.
Periods of very intense photon activity are selected through  
a novel analysis approach. 
The statistical method called {\it Maximum Likelihood Blocks} 
is applied for the first time on light curves of high frequency blazars. 
In order to avoid any possible bias in the selection of periods with intense photon activity, 
the arrival time and incoming direction of the neutrinos are kept blinded.
Following the approach here reported, neutrino fluxes 
below the atmospheric neutrino background level can become accessible.
We report as well on a first step to establish a target-of-opportunity program 
based on neutrinos detected in IceCube which are used as alerting messenger particles. 
\end{abstract}

\section{Introduction}
We report here on the inclusion of the photon flux time evolution, measured in one or more photon
wavebands, in the search for HE neutrino sources.
We illustrate this approach by a discussion of blazars. 
In the framework of  {\it hadronic models} \cite{K. Mannheim, Anita}, 
blazars are HE neutrino sources and they
are dominated by a highly variable component of non-thermal 
radiation \cite{Begelman}. 
\noindent
Blazar broad-band spectra consist of two prevalent components which appear in the 
spectral energy distribution like two broad humps. 
The low-energy component can peak at various frequencies between optical and  X-ray 
and the high-energy is proportionately shifted from X-rays up to very high energy (VHE) $\gamma$-rays. 
\noindent
 The synchrotron radiation coming
 from primary electrons as well as  from electrons produced in proton-induced 
 cascades contributes to the low-energy component.
 In hadronic models, high-energy radiation arises  from
 photo-meson interaction and from proton and muon synchrotron radiation \cite{Anita}.  
The $\gamma$-ray production by pion photo-production is accompanied by neutrinos, 
created in the decay of charged pions.\\

\noindent
So, assuming that neutrino production follows  
the same time behaviour like the electromagnetic activity, 
the transient nature of the energetic emission can be used to improve the association between 
highly energetic neutrinos and non-thermal sources. 
If the enhancement of the neutrino signal is concentrated in a short period of time, 
 neutrino flares not evident in a time-integrated point source search 
like the one reported in \cite{PS}, might be detectable.  
A first analysis following this approach has been discussed in \cite{ICRC, EB}. 
In this paper, we concentrate on the statistical interpretation of measured
light curves.

\section{Collection and Interpretation of Light Curves: Periods Selection}
\noindent
IceCube Collaboration has started a comprehensive collection of 
historical light curves. For now, most of our efforts are concentrated on X- and VHE $\gamma$-ray wavebands
which corresponds to the two humps of the spectral energy distribution for high frequency blazar (HBL).
The two instruments used for X-ray data are All-Sky Monitor (ASM) \cite{ASM} and 
the Proportional Counter Array (PCA) installed on board of the Rossi X-ray Timing Explorer (RXTE).
RXTE standard data products are collected directly from the HEASARC database and then transformed in
{\it root} format.  
A large set of data has been collected by VHE $\gamma$-ray experiments \cite{Martin}. 
Moreover, optical data has been as well taken for few sources and reported in \cite{Teresa}.\\

\noindent
In order to utilize the transient character of the electromagnetic emission for HE neutrino search,
we have first to separate 
variable  periods ({\it flares}) and steady state periods of a source.
Periods of no variable activity are often defined in the literature as {\it quiescent} 
but an apparent {\it quiescent} level can be due to 
a superposition of numerous unresolved flares or to a limited sensitivity of
the instrument \cite{wolk05}.  We call the level of activity in which the source stays for the longest period of time its
{\it characteristic} level. We do not attempt to interpret this level 
in phenomenological terms.
Often data are affected by large uncertainties or the data spacing is rather inhomogeneous.
In order to improve the interpretation of photon data, 
a simple and model-independent approach has been applied. 
The method aims at dividing the light curves in time
intervals in which the source emission is compatible with a constant
level. 
An algorithm based on Bayesian statistics that provides such a
segmentation of data of different nature was presented in
\cite{scar98} and a modified version based on
Maximum Likelihood was recently employed in studies of stellar
X-ray light curves \cite{wolk05}. We will refer to this
algorithm as the method of Maximum Likelihood Blocks (MLBs); its first application
to blazars X- and VHE  $\gamma$-ray light curves has been reported in \cite{internal}. \\

\begin{figure}[ht]
\begin{center}
\includegraphics*[width=0.8\textwidth,angle=0,clip]{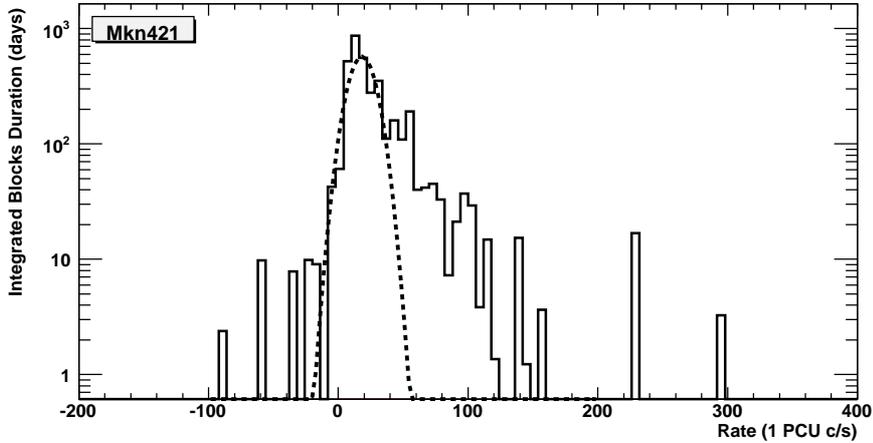}
\caption{Duration of maximum likelihood blocks for Mkn421, based on ASM flux.  
This represents the measurement of the integrated time a source stays at a given flux level.
In particular, the flux of the peak is the flux level in which the source stays for the longest 
time and therefore corresponds to the characteristic level. 
Negative blocks are due to systematic errors in ASM. }  
\label{gauss421}
\end{center}
\end{figure}

\begin{figure}[ht]
\begin{center}
\includegraphics*[width=0.8\textwidth,angle=0,clip]{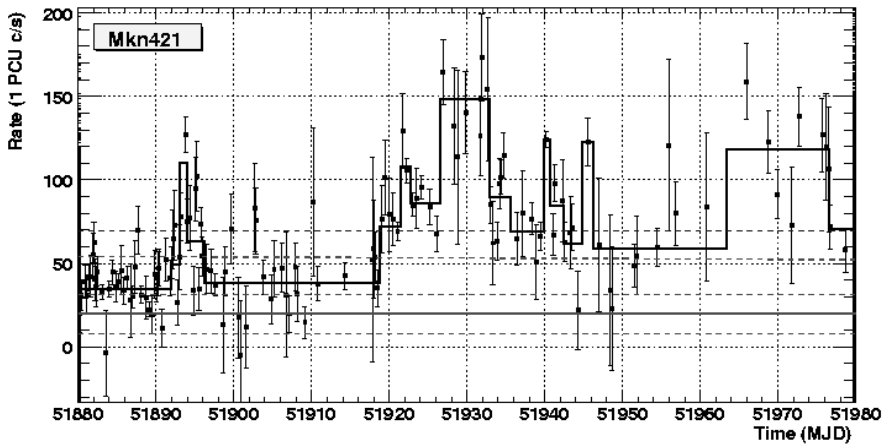}
\caption{\label{fig.1}  Mkn421, sub-period of 100 days of ASM light curve (10 years total time of observation), 
blocks (99\% C.L.). The continuous red line corresponds to the {\it characteristic level} $R_{char}$, the dotted line
represent 1 $\sigma_{char}$ and 3 $\sigma_{char}$ deviation from $R_{char}$.
The  blue line corresponds to $R_{char} + 5 \sigma_{char}$.
The X-ray units are normalized to 1 PCU count/second 
(1 Crab =$\sim$ 3000 counts/sec/detector).} 
\label{2}
\end{center}
\end{figure}

\noindent
The MLBs sub-divides the light curve into constant-flux intervals 
or {\it blocks}. The confidence level at which the  
algorithm splits the light curve is given as input to the algorithm (in this work 99\%).
From this interpretation of the data we can extract various information about the source like
the time a source pass in a
particular activity state, if there are favorite flux levels and the periods in which the
source is in {\it flare} state. For a visualization, 
the flux value of the single block is histogrammed and the duration of the block is used 
as a weight for the single entry.
In Fig.~\ref{gauss421}, the histogram of ASM blocks are reported for the Mkn421 as example.   
The peak of the distribution is quite naturally interpreted as the {\it characteristic} level. 
The tail at higher flux values represent the 
flaring activity of the source; the tail at negative flux  indicates measurement errors 
 in the ASM data. 
Mean value and the standard deviation
of a fitted Gaussian are then interpreted as the {\it characteristic} level $R_{char}$ and 
$\sigma_{char}$, respectively. 
An example of such interpretation is shown in Fig.~\ref{2}. 
With this interpretation of the light curve,
the periods of time when the source is in {\it flare} can be selected requiring that
the photon flux deviates from the characteristic level for a certain number of 
standard deviations. In this way, a uniform selection of {\it flares} is obtained  on years-long light curves.
A similar procedure applied on VHE data failed in the identification of a possible VHE $\gamma$-ray characteristic
level. Nevertheless, an arbitrary flux threshold can be placed in order to select flaring periods on VHE light curves. 
Considerations about the VHE data are reported also in \cite{Martin, Teresa}.\\

\noindent
A comparison of the time behaviour of different wave length bands may lead to a particularly
interesting classification of periods for the neutrino search, as will be described below.
Nearly simultaneous X- and VHE $\gamma$-ray data are collected from organized multi-wavelength campaigns.
The time and flux correlation between the X- and $\gamma$-ray 
components have been investigated 
\cite{Krawczynski501_2000}.
In most of the cases, a time-correlation between the two components has been observed as expected
in the framework of {\it leptonic models}.
So called orphan flares, 
where an increasing TeV flux is not accompanied by an increase of the
X-ray flux,  have been seen in at least two
sources: 1ES~1959+650 and Mkn~421. 
A study of the correlation between these two wavebands using the high statistic light curves
collected is under way and results will be reported elsewhere.

\section{Toward a Multi-messenger Approach}
The use of historical light curve to improve the search
for HE neutrino sources is limited by  the low duty cycle of $\gamma$-ray telescopes and rare long term observations. 
Neutrino telescopes, on the contrary, like IceCube are characterized by a very wide field of view and
very high duty cycle. 
HE neutrinos are produced exclusively by hadronic mechanisms and are expected to be time-correlated
 with the related photon emission.
In order to improve the synchronous measurement of both photons and neutrinos,  
the idea to develop a {\it hadronic} trigger or target of opportunity (ToO)
using HE neutrino candidates has been born within the IceCube Collaboration \cite{EB}.
The ToO under discussion concerns  
transient phenomena of particularly interesting targets, 
as well as unpredictable sudden astronomical events.
HE neutrino candidates are reconstructed on-line at the South Pole as up-going muon tracks. 
Information  on these events can be transferred north via satellite and, in principle, be used
as the alert messengers for other telescopes.
The technical realization of such alerts is under investigation and results are encouraging.
The first telescope already reacting to neutrino ToO program  is the MAGIC VHE $\gamma$-ray telescope. 
A test run started on September 27 (2006) is based on AMANDA data acquisition system and will last for a few months. The test 
is focused on technical aspects, such as the recording and real time reconstruction of the neutrinos 
at the South Pole, the reception of the triggers by MAGIC and the communication between the two instruments.\\

\noindent
To date, no indication of HE cosmic neutrinos has been found in the performed analysis of AMANDA data. 
Contrary to other ToO programs, the alert in the case of the neutrino ToO is issued on the basis
of a non clear detection.
Non-trivial statistical issues related to the interpretation of possible coincidences
are under careful investigation and will be reported elsewhere. 
Beyond that, once a significan accumulation of HE neutrinos will be observed in IceCube, 
 the simultaneous photon data provided by ToO programs will lead to a mature phenomenological picture
 of the astronomical object observed.\\

\newpage

\title{Air showers in a three dimensional array: Recent data from IceCube/IceTop
\footnote{Research supported by the U.S. National Science Foundation}
}
\author{Xinhua Bai \& Thomas K. Gaisser for the IceCube Collaboration} 
\address{
Bartol Research Institute and Department of Physics and Astronomy\\ 
University of Delaware,
Newark, DE 19716 USA}

\ead{bai@bartol.udel.edu, tgaisser@bartol.udel.edu}

%\maketitle
\begin{abstract}
The next generation high energy neutrino and cosmic ray array 
IceCube/IceTop is under construction at the geographic South-Pole. 
Air showers with trajectories that pass through the surface array 
and near the deep strings trigger both components in coincidence. 
The ratio of the muon signal in the deep detectors to the shower 
signal on the surface is sensitive to the elemental composition of 
the primary cosmic radiation. 
\end{abstract}

\vspace{-1pc}
One string of 60 sensors buried between 1.5 and 2.5 km in the ice and a 
surface array of 4 stations were successfully deployed at the 
South Pole during the austral summer of 2004-05 and have been producing 
data since February 2005~\cite{performance05}. Eight more strings and 12 more 
IceTop stations were deployed in the austral summer of 2005-06.  Since then 16 stations 
and 9 strings have been operating. The full array with up to 80 strings
and 80 surface stations is scheduled for completion in 2011.

Each IceTop station consists of a pair of ice Cherenkov tanks (to be referred 
to as tank A and B) separated by $10$~m, each containing a cylinder of clear 
ice $2.7$\,m$^2\times 0.9$\,m viewed from the top by two standard IceCube digital optical modules (DOMs).
%The IceTop DOMs are fully integrated into the calibration and data acquisition system
%of IceCube. 
The operation of a surface array over the deep IceCube neutrino 
telescope has three goals:
             \begin{itemize}
             \item { Composition: To study the ratio of the muon signal in the deep array
to the shower signal on the surface which is sensitive to the fraction of heavy nuclei 
in the cosmic-ray spectrum.} 
             \item { Calibration: To study the angular resolution and 
pointing accuracy of the neutrino array by providing a sample of externally 
identified muon bundles.}  
             \item { Filtering: To study and filter single or multiple 
muon background in the deep detector by tagging the associated air-shower 
activities on the surface.}
             \end{itemize}

Predecessors for operation of a surface array in conjunction with a deep underground
detector sensitive to muons are SPASE-AMANDA~\cite{composition} and
EASTOP-MACRO~\cite{EASTOP-MACRO}.  With an acceptance at completion of 
$\sim 0.3$~km$^2$sr, IceTop/IceCube will have a significantly higher reach in
primary energy than these earlier experiments.  With a threshold of approximately
$300$~TeV, the experiment will be sensitive from below the knee of the
cosmic-ray spectrum up to approximately $1$~EeV, where it will be statistically
limited by its acceptance.  A significant motivation for studying the composition
in this energy region is to search for the transition from a population of
cosmic rays primarily of local origin in the Milky Way Galaxy to a population
of extra-galactic origin~\cite{hillas06}.  

\begin{figure}[h]
\vspace{0pc}
\hspace{2pc}\begin{minipage}{16pc}
\begin{center}
\includegraphics[width=16pc,height=13pc]{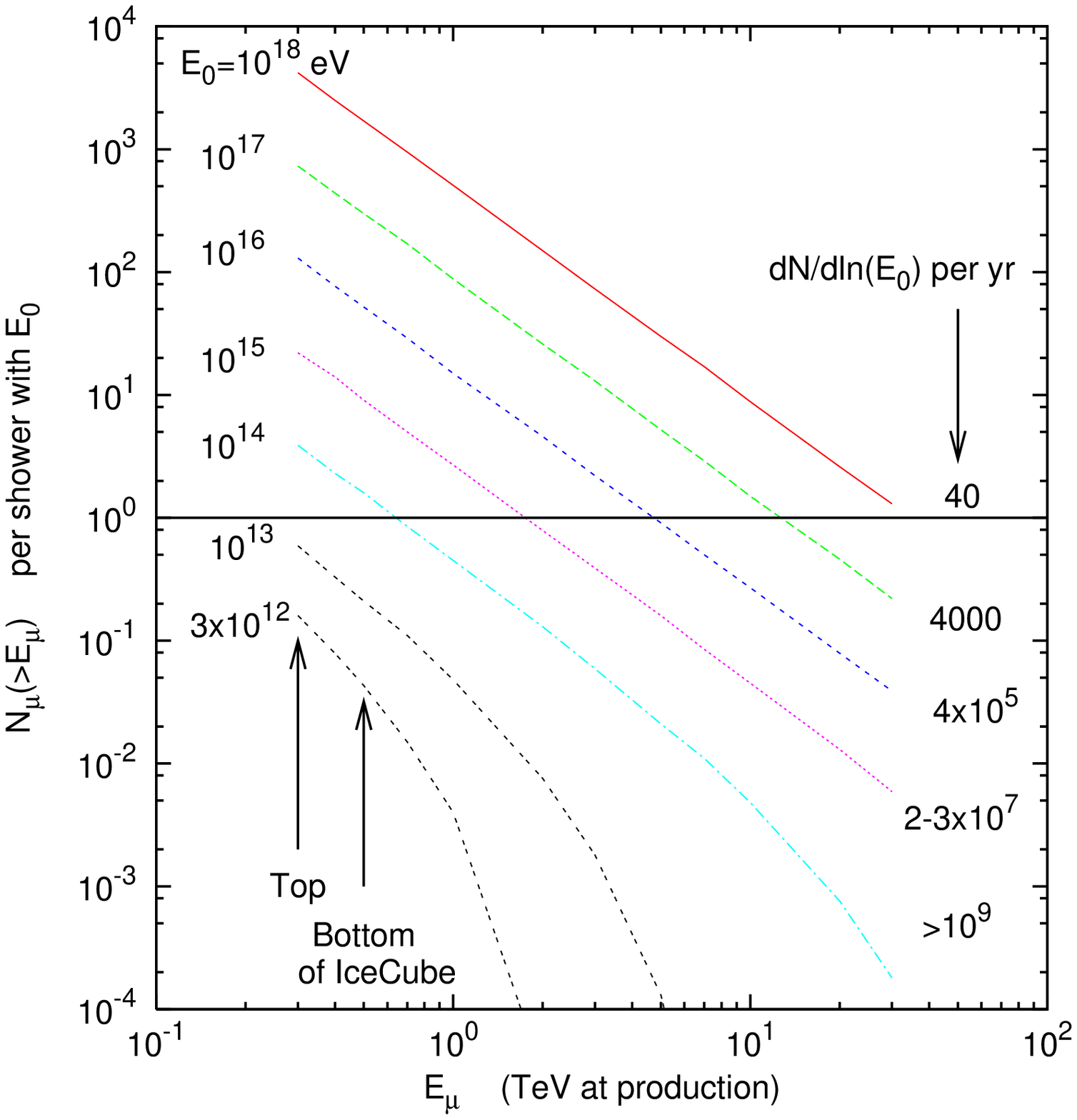}\vspace{-1pc}
\caption{\label{fig1}Integral energy spectra of muons in air showers (see text).}
\end{center}
\end{minipage}\hspace{1pc}%
\vspace{-1pc}\begin{minipage}{16pc}
\begin{center}
\includegraphics[width=16pc,height=13pc]{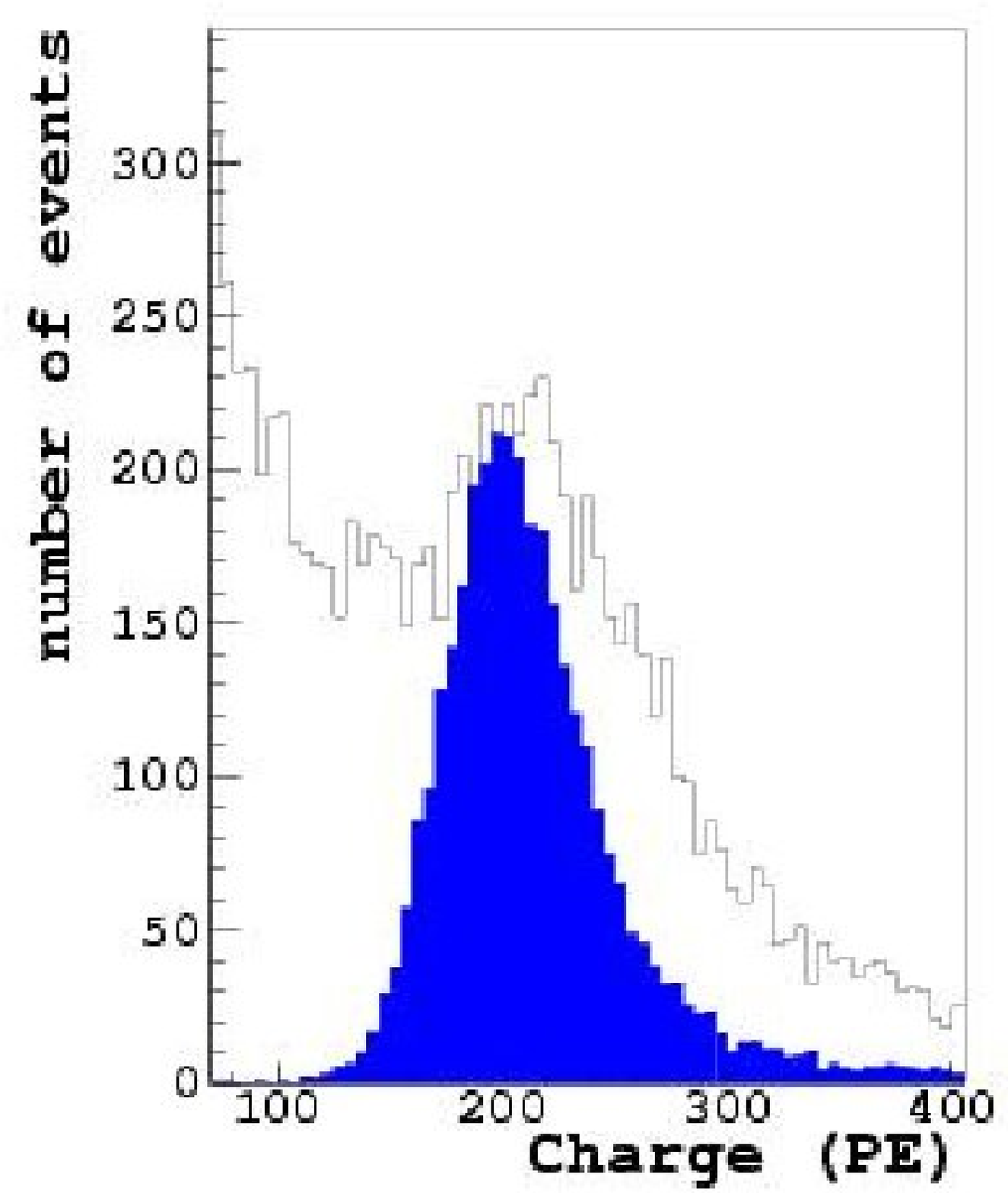}\vspace{-1pc}
\caption{\label{fig2a}Charge spectrum of pulses in an IceTop tank (see text).}
\end{center}
\end{minipage} 
\end{figure}
\vspace{1pc}

Figure~\ref{fig1} shows the integral energy spectra of muons at production in the
atmosphere in proton-initiated showers of various primary energies.
The two vertical arrows indicate the minimum muon energies needed to reach the top
and bottom of the deep in-ice detector.  Each spectrum is 
labeled on the right with the number of coincident events
per year (d$N$/d\,ln$E$)  expected within the acceptance of the full detector.  
The currently operating array with
16 surface stations and 9 strings has approximately 0.5\% of the geometrical 
acceptance for coincident events of the full detector and is therefore
statistically limited at $10^{17}$~eV, where the expected number
of events per year would be about $20$ (as compared to $4000$ for the completed detector).

Low-energy atmospheric muons (typically in the GeV range) provide a natural beam
for calibrating and monitoring the response of IceTop
detectors to track length above Cherenkov threshold and hence to the
energy deposition in the tanks.  The characteristic spectrum (shown 
in Fig.~\ref{fig2a})
combines the steeply falling spectrum of electrons and converting $\gamma$-rays with a peak 
due to muons. Small air-showers contribute to the 
high-energy tail.  The solid histogram in Fig.~\ref{fig2a} shows
the single muon peak identified by a muon telescope in a special run.
The tagged muon histogram is narrower than the muon peak in the
composite spectrum and very slightly shifted toward the lower integrated charge.
The vertical through going muon deposits $160$~MeV and thus provides
the conversion between energy deposition and integrated charge of the waveform.
Special, periodic monitoring runs obtain the composite, inclusive spectrum to look for
any change in shape or peak location, which would indicate a change in tank response.

In normal data taking, the IceCube data acquisition system sends data to 
the surface only if neighboring DOM pairs are hit. For IceTop, we require 
that both tanks at the same station register a hit so that only air showers 
are reported.
Events are recorded if the hits satisfy a simple majority trigger (SMT).
For IceTop, the SMT is set to $6$ DOM hits within $2 \mu s$; for in-ice 
detector, the SMT is set to $8$ hits within $5 \mu s$.  With these settings, the
in-ice trigger rate is $146$~Hz and the IceTop SMT rate is $7.1$~Hz.
Whenever either trigger is satisfied, waveforms of all hit DOMs
are recorded.  Of particular interest is the subset in which
both SMT triggers are satisfied.  This coincident rate is measured to be 
$0.19$~Hz in the 2006 detector configuration. These events will be the subject
of the composition analysis. They also provide tagged muon beams for 
the calibration of in-ice array.  

\begin{figure}[h]
\vspace{0pc}
\hspace{2pc}\begin{minipage}{16pc}
\begin{center}
\includegraphics[width=16pc,height=14pc]{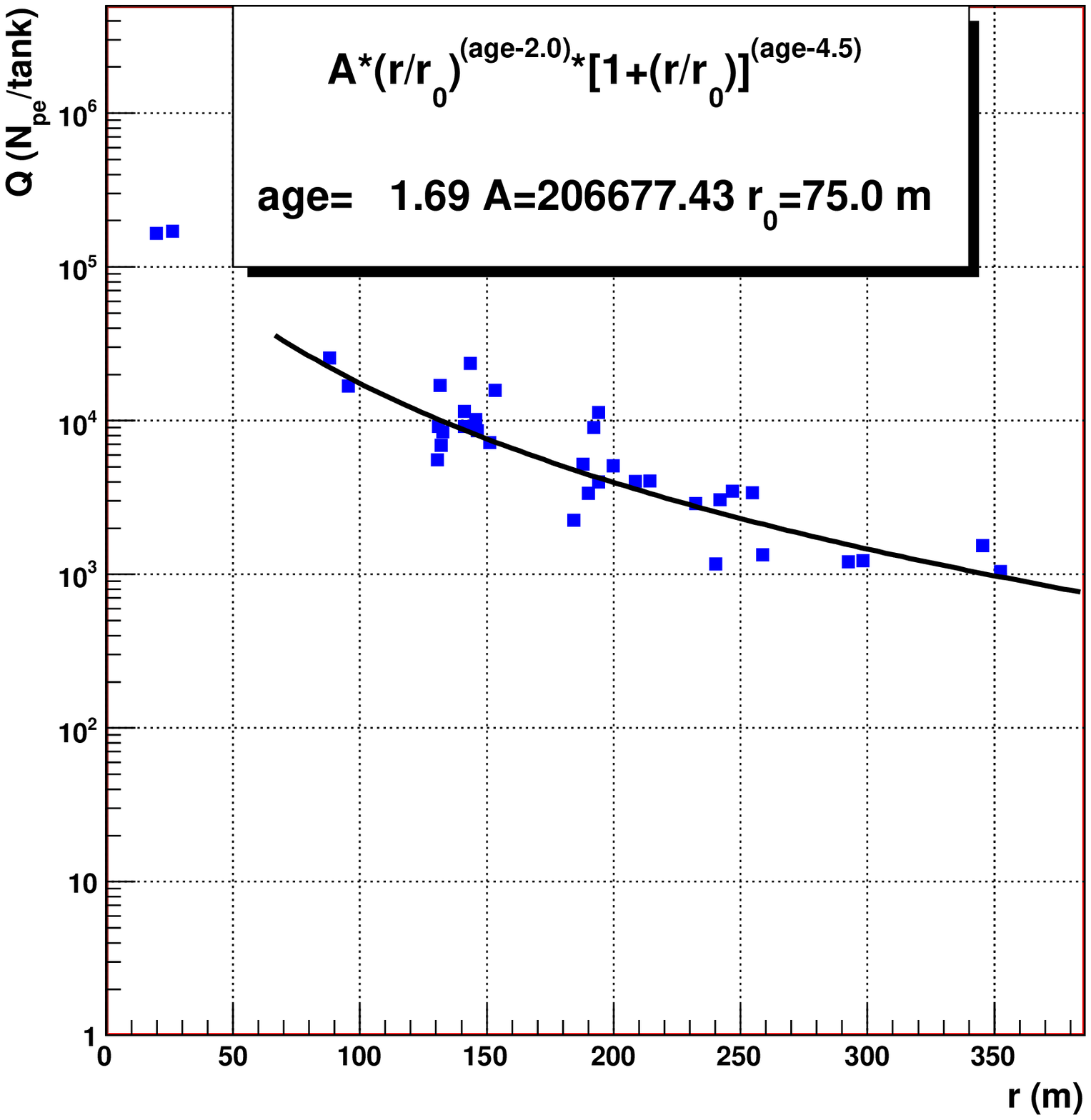}
\caption{\label{fig4a}Lateral distribution of the charge in surface
tanks fitted with an NKG distribution.}
\end{center}
\end{minipage}
\hspace{1pc}%
\begin{minipage}{16pc}\vspace{-1pc}
\begin{center}
\includegraphics[width=15pc,height=14pc]{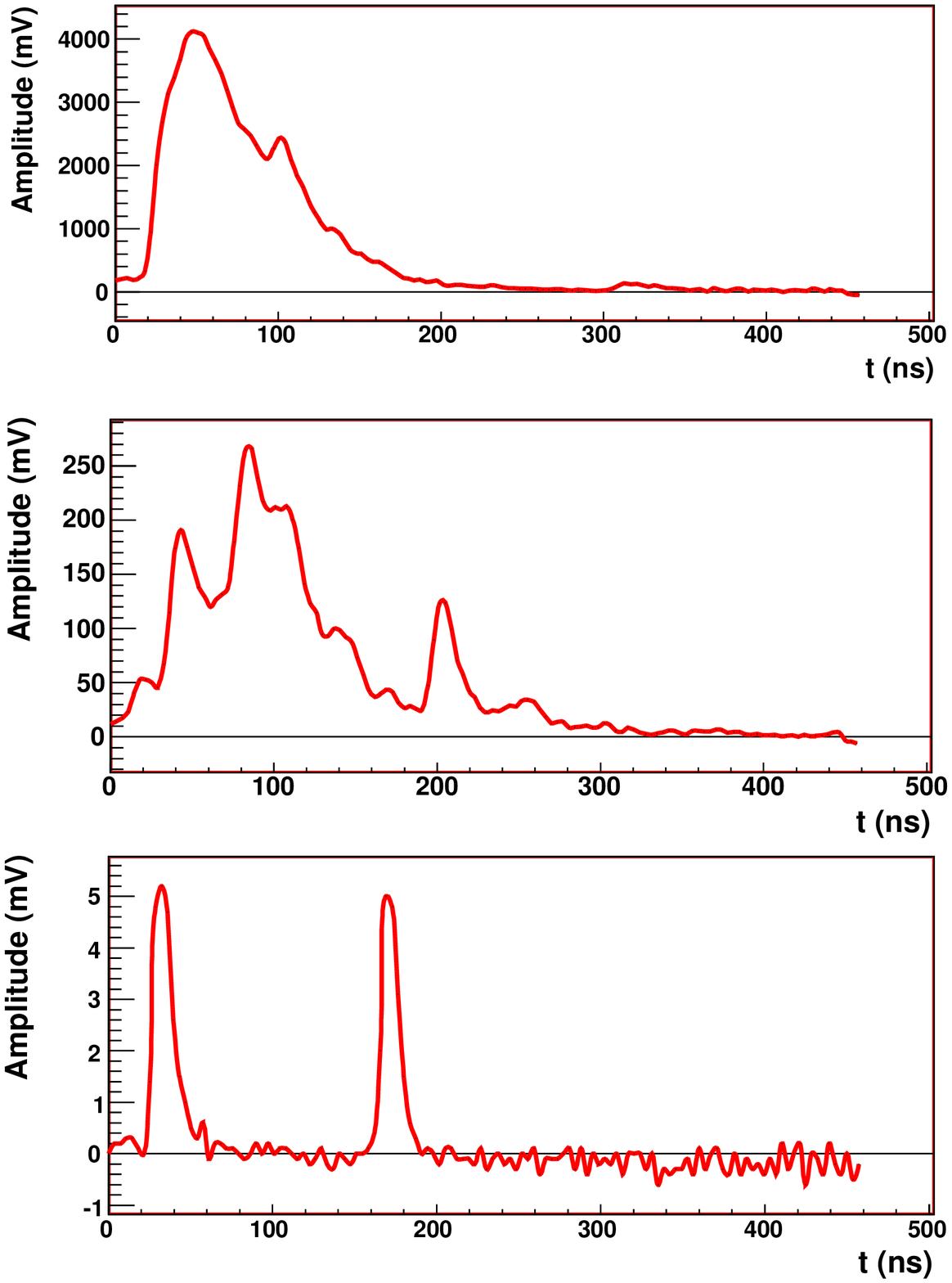}
\caption{\label{fig4b}  Waveforms in coincident events (see text).}
\end{center}
\end{minipage} 
\end{figure}
\vspace{0pc}

Figure~\ref{fig4a} shows the lateral distribution for a typical large
IceTop event, which happens to be a coincident trigger. Figure~\ref{fig4b}
shows the waveforms on the surface at two locations ($\sim100$~m and $\sim210$~m 
from the reconstructed shower core) 
and a sample waveform from an in-ice DOM in the same event.  Surface
waveforms have the characteristic features 
of large, smooth shape near the core and smaller, more uneven structure 
farther out. In-ice waveforms are typically a sequence of
single photo-electrons.

\begin{figure}[h]
\vspace{0pc}
\hspace{2pc}\begin{minipage}{15.5pc}
\begin{center}
\includegraphics[width=16pc,height=11pc]{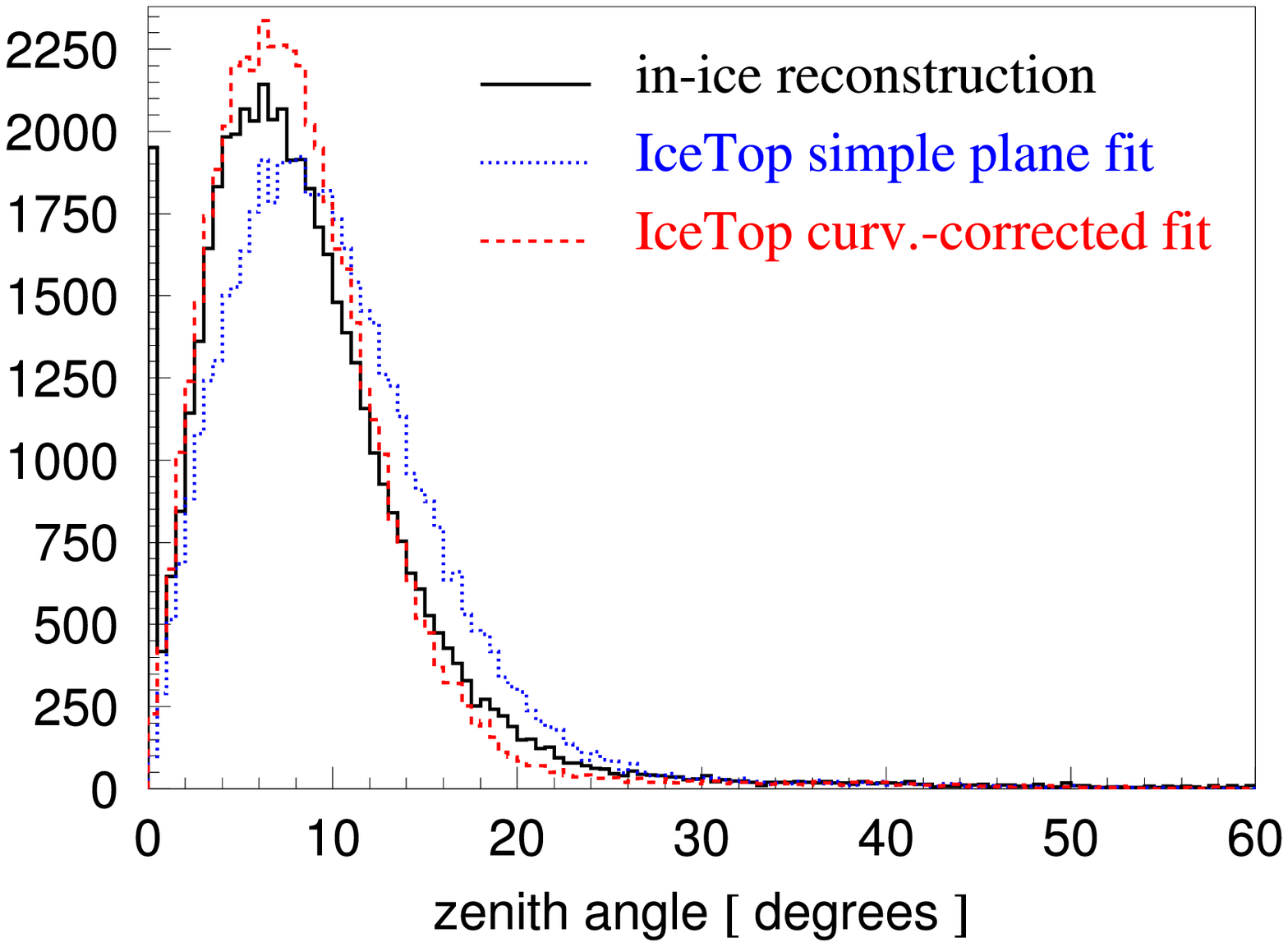}
\caption{\label{fig5a}Zenith angle distribution of coincident
events as reconstructed by in-ice alone (black histogram)
and by IceTop alone with two different procedures.}
\end{center}
\end{minipage}\hspace{2pc}%
\begin{minipage}{15.5pc}
\begin{center}\vspace{-1pc}
\includegraphics[width=16pc,height=11pc]{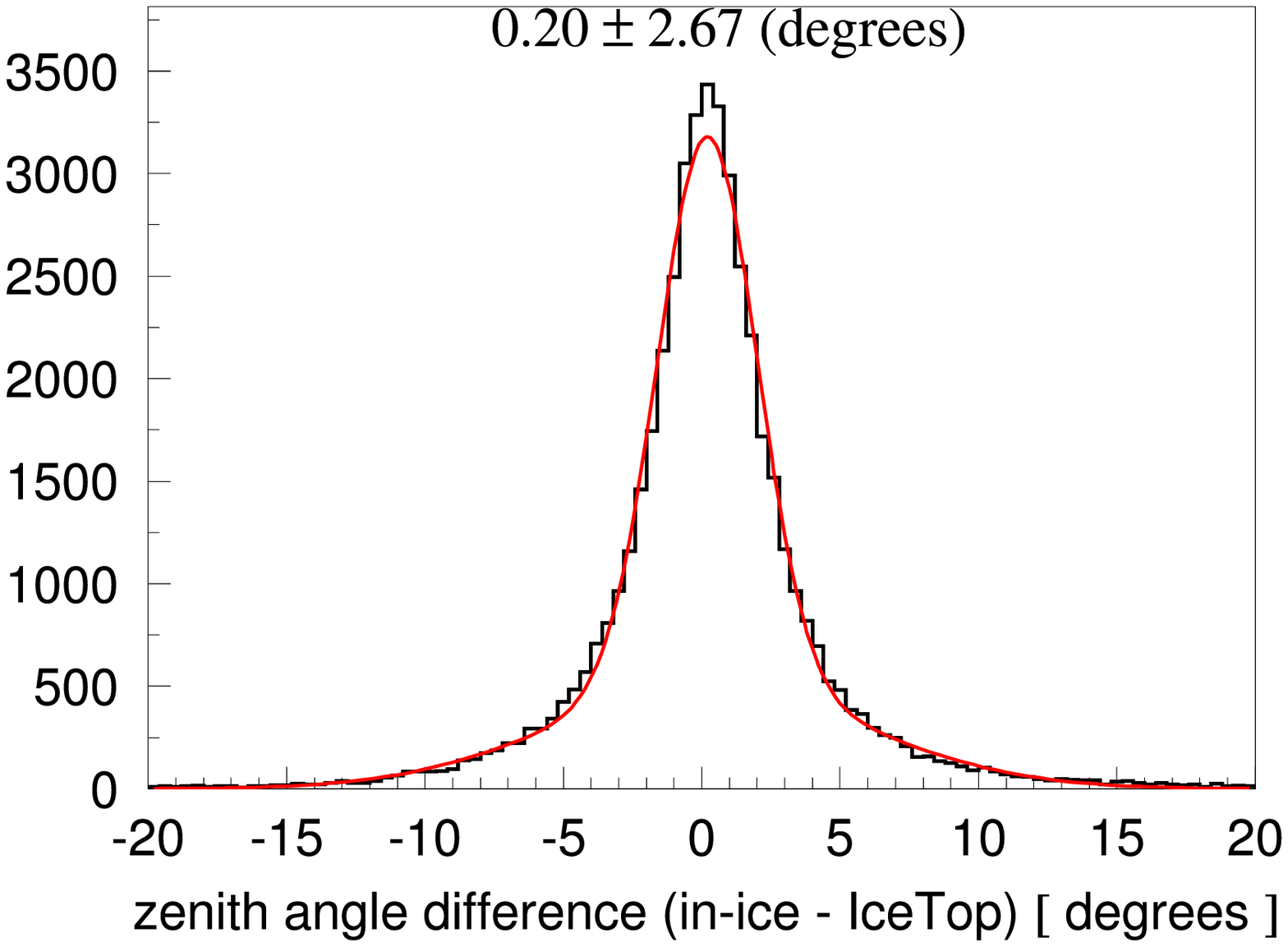}
\caption{\label{fig5b}  Distribution of the difference between directions
determined by IceTop alone and by using the in-ice muon reconstruction algorithm.}
\end{center}
\end{minipage} 
\end{figure}

\vspace{0pc}
Finally, Figs.~\ref{fig5a} and \ref{fig5b} illustrate cross calibration
of angular resolution between IceTop and the deep-ice array of IceCube.
When the realistic curved shower front (dashed in ~\ref{fig5a}) is used, the directions agree
well (FWHM$\approx 5^o$).  A sub-array analysis is in progress to determine
the angular resolution of the IceTop reconstruction algorithm.  This analysis
uses the pairwise distribution of tanks to form one sub-array of all 16 ``A"
tanks and an second sub-array of all 16 ``B" tanks at each station.
Comparison of the separately determined ``A" and ``B" directions for each
event give a measure of the resolution of IceTop alone.  Deconvolving
the distribution of Fig.~\ref{fig5b} will then give a measure
of the resolution of the in-ice reconstruction algorithm as applied to
muon bundles.

\newpage
\setcounter{section}{0}
\title{High-Energy Gammas from the giant flare of SGR 1806-20 of December 2004 in AMANDA}

\author{Juan-de-Dios Zornoza for the IceCube
Collaboration\footnote{http://www.icecube.wisc.edu}}

\address{222 W. Washington Ave. Madison WI 53703 (USA)}

\ead{zornoza@icecube.wisc.edu}

\begin{abstract}
We show in this paper the analysis of the AMANDA-II data looking for
events correlated with the giant flare observed in December 27th 2004 from
the Soft Gamma-ray Repeater 1806-20. This flare was more than two orders
of magnitude brighter than any previous flare of this kind and
saturated the satellite gamma detectors that observed it. If a hard
component of gamma-rays was present in the event, these would produce
detectable rates of muons in underground detectors like
AMANDA. Moreover, high-energy neutrinos could also have been emitted in
quantities large enough to produce a signal in this detector. The
unblinding of the data showed no signal, so upper limits were set both
to the gamma-ray and the neutrino fluxes.
\end{abstract}

\section{Introduction}

Soft Gamma-ray Repeaters (SGRs) are X-ray pulsars which emit X-ray
bursts lasting $\sim$0.1~s during sporadic active periods. The typical
luminosities of these bursts is $10^{41}$~erg/s. However, there are
rare occasions in which giant flares (in X-rays and soft-gamma rays)
are observed, with luminosities thousands of times higher than normal
bursts. Three of these giant flares had been observed until
1998~\cite{sgr1}. On December 27th 2004, a giant flare of soft-gamma
rays and hard X-rays coming from the Soft Gamma-ray Repeater 1806-20
saturated several satellite
gamma-detectors~\cite{integral,swift-bat,rhessi}. This was the
brightest transient event ever observed in the Galaxy.

The most accepted theory to describe SGRs is the ``magnetar''
model. According to this model, these objects are very-rapidly-rotating neutron
stars, with extremely high magnetic fields ($B\sim 10^{15}$~G, two
orders of magnitude larger than in normal neutron stars). Along time
scales of the order of tens of years, these strong magnetic fields
build up an increasing stress in the star. When the stress on the star
crust is too strong, it fractures. This produces a starquake which
liberates enormous quantities of energy in X-rays and $\gamma$-rays
as the magnetic field rearranges~\cite{Kouveliotou}.

The energy spectrum of the Dec. 2004 flare can be described as the sum
of a black-body spectrum and a power law~\cite{Palmer}, which would
indicate a relevant component of high-energy ($\sim$TeV)
emission. High-energy gammas would produce showers when interacting at
the top of the Earth's atmosphere. These showers would be muon-poor,
but some of the many photons produced in these interactions would
produce pions. The decay of these pions would yield muons which can
reach underground detectors like AMANDA~\cite{ourtheory}.

High-energy neutrino fluxes have been also predicted by some
authors~\cite{Gelfand, Ioka}, if there is a significant
baryonic outflow. In this case, neutrinos could also reach underground
neutrino detectors and produce a signal.

\section{The AMANDA detector}

The AMANDA neutrino telescope~\cite{Andres} consists of a three
dimensional array of 677 Optical Modules (OMs). An Optical Module is
basically a photomultiplier and its electronics housed in
pressure-resistant glass sphere. These OMs are distributed along 19
strings buried 1500-2000~m deep in the Antarctic ice.

The main aim of this experiment is the detection of cosmic neutrinos.
The principal signature is given by high-energy neutrinos interacting
in the surroundings of the detector and producing a relativistic muon
which would emit Cherenkov light when traveling in the ice.  Events
are recorded when at least 24 OMs register a signal within
2~$\mu$s. The information of the position and the time of the photons
hitting the photomultipliers is used to reconstruct the direction of
the neutrino. As we have mentioned before, the main motivation in this
analysis is the search for muons produced indirectly in the showers
induced by gamma-rays interacting in the top atmosphere. Since the
source was above the horizon, the effective area for TeV photons is
one order of magnitude higher than for neutrinos. However, there is no
way to distinguish between both possibilities in case of the
observation of a muon.

\section{Analysis}

Since both the time and the position of the burst can be well
constrained, the enormous background of muons produced by cosmic rays
in the atmosphere can be effectively reduced. 

There are two variables to optimize in this analysis: the width of the
time window and the angular size of the search cone. In order to
prevent a possible bias in the analysis, we perform the optimization
of the selection criteria with the data blinded, which is particularly
relevant when small signals are expected, as it is our case.

The duration of the burst was $\sim$0.6~s. However, this window had to
be widened in order to account for the dispersion in the times given
by the satellites~\cite{geotail,private,integral,rhessi,cluster4},
calculated at the location of the detector. The chosen window, once
these facts were taken into account is 1.5~s around UT 21h 30m 26.6s
of December 27th.

The next step is to determine the best search cone. This is done by
optimizing the so-called Model Discovery Factor (MDF)~\cite{hill}, defined as

   \begin{equation}
     \label{eq:mdf}
     MDF=\frac{\mu(n_b, CL, SP)}{n_s}
   \end{equation}

\noindent where $\mu$ is the Poisson mean of the number of signal
   events which would result in rejection of the background
   hypothesis, at the chosen confidence level $CL$, in $SP$\% of
   equivalent measurements. $n_s$ and $n_b$ are the number of signal
   and background events, respectively.

The background was determined using on-source, off-time, real data
(the data $\pm$10 minutes around the burst is kept blinded). It was also
checked that the detector rate on December 27th was stable (rate$\sim$90~Hz,
close to the AMANDA average).

The signal is simulated in order to estimate the angular resolution
and the effective area of the detector. With the codes
CORSIKA-QGSJET01~\cite{corsika} and ANIS~\cite{anis}, we generated
photons and neutrinos, respectively, with energies ranging from 10~TeV
to 10$^5$~TeV. The secondary muons are propagated up to the detector
and reconstructed.

\begin{figure}
 \begin{center}
\begin{tabular}{cc}
\includegraphics[width=0.5\linewidth,angle=0]{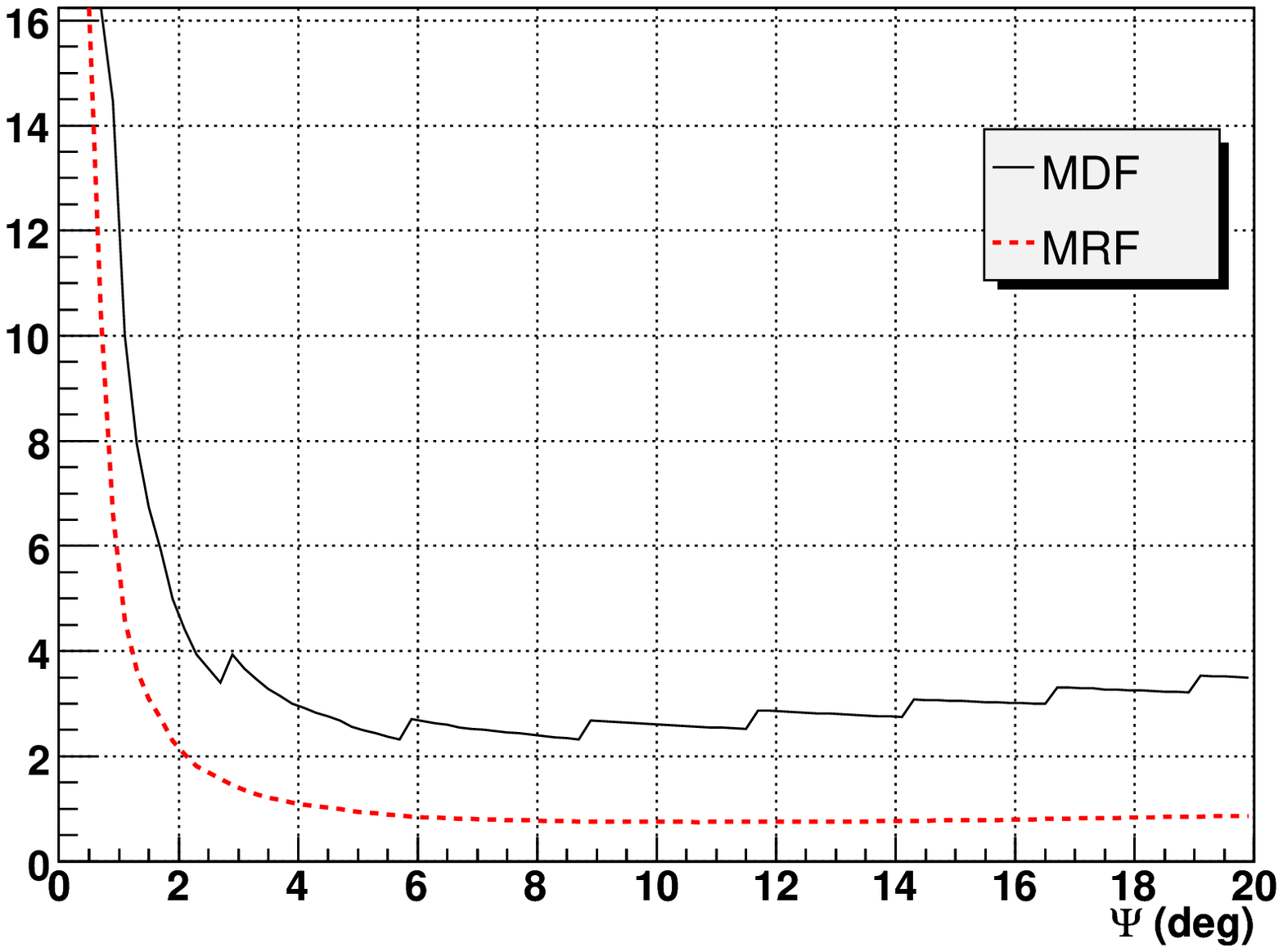} &
\includegraphics[width=0.5\linewidth,angle=0]{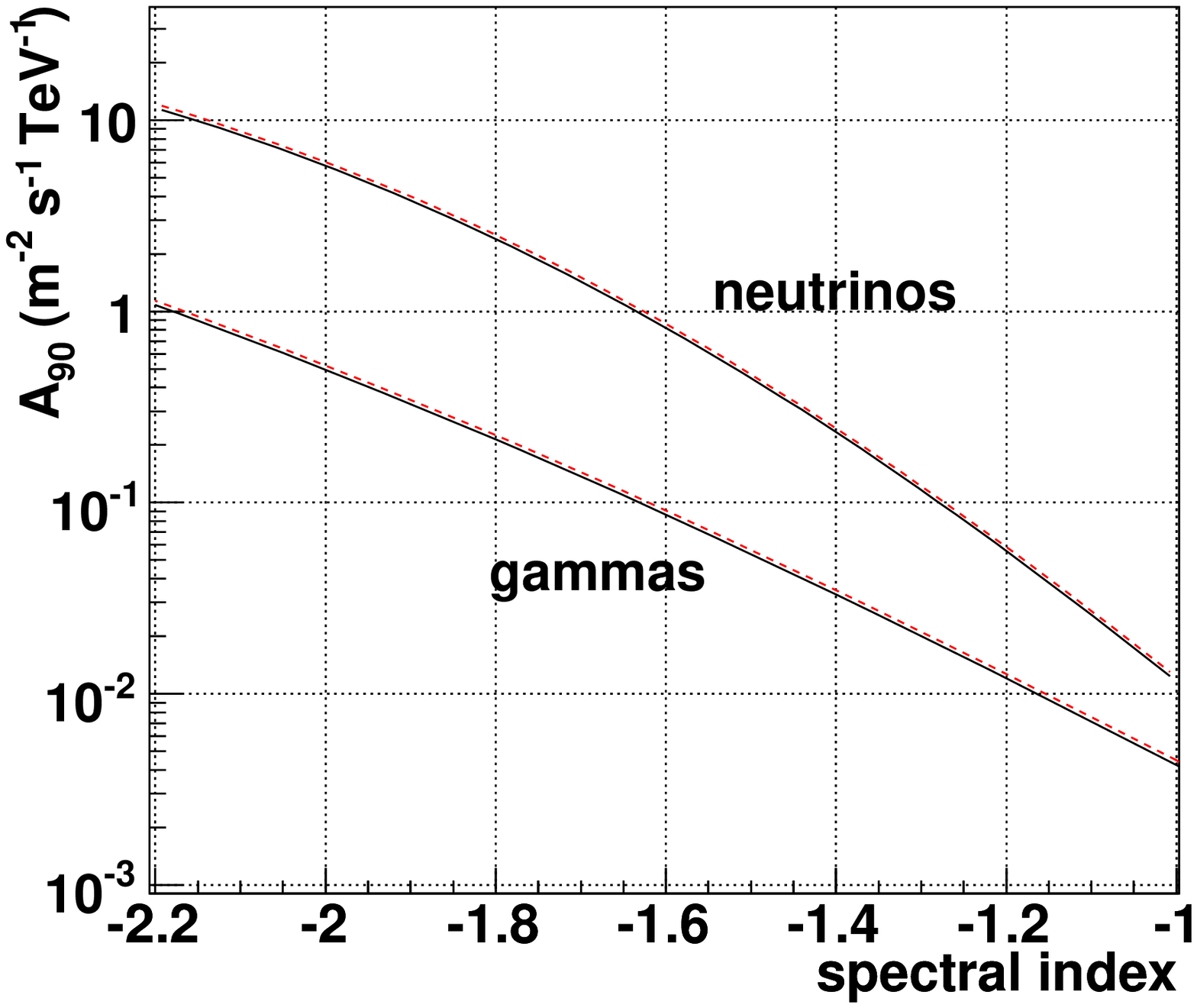}
\end{tabular}
 \end{center}
 \caption{Left: Model Discovery Factor (solid, black) and Model
 Rejection Factor (dashed, red) as a function of the radius of a
 circular angular acceptance window, for an $E^{-1.47}$
 spectrum. Right: Sensitivity (dashed, red) and limit (solid, black)
 to the normalization constant in the flux of gammas (lower line) and
 neutrinos (upper line), as a function of the spectral index, assuming
 a flux $\phi(E)=A~(E/{\rm TeV})^{\gamma}$.}
 \label{plots}
\end{figure}

The dependence of the MDR on the search cone is shown in
figure~\ref{plots} (left). It can be seen in that plot that the
optimum size corresponds to a radius of 5.8$^\circ$. The expected
background for such a cone, during 1.5~s is 0.06 events, at the
location of the source.

% ------------------------------------------------------------------------
% ------------------------------------------------------------------------

\section{Results}

Once the optimum selection criteria were found, the data were
unblinded. However, no event was found correlated with the
burst. Therefore, upper limits were set, based on the effective area
of the detector. Assuming a power law spectrum $\frac{dN}{dE} < A_{90}
(E/\mathrm{TeV})^{\gamma}$, the limits on the normalization constant
of the flux of gamma-rays and neutrinos are shown in
figure~\ref{plots} (right), as a function of the spectral index. This
means, for instance, limits of 0.05 (0.5) TeV$^{-1}$~m$^{-2}$~s$^{-1}$
for $\gamma=-1.47$ ($-2$) in the gamma flux and $0.4$ ($6.1$)
TeV$^{-1}$ m$^{-2}$ s$^{-1}$ for $\gamma=-1.47$ ($-2$) in the
high-energy neutrino flux (at 90\% CL).

\section*{Acknowledgments}

This work has been done with the financial support of the Marie Curie
OIF Program.

\newpage
\setcounter{section}{0}

\title{IceCube - First Results}

\author{Jon Dumm and Hagar Landsman for the IceCube Collaboration}

\address{Department of Physics,University of Wisconsin, Madison, Wisconsin, USA}

\ead{hagar@icecube.wisc.edu, jdumm@icecube.wisc.edu}

\begin{abstract}
 During the last two austral summers, the first sensors of the IceCube neutrino observatory were deployed in the deep Antarctic ice, along with a surface array. We will present first results obtained using the IceCube detector, demonstrating that the performance is within the design requirements, and showing the ability to reconstruct tracks, cascades and synchronizing times in the entire array to within 3 ns.
\end{abstract}

\section{Introduction}
The IceCube detector currently under construction at the South Pole, 
will consist of up to 4800 Digital Optical Modules (DOMs) covering a 
fiducial volume of 1 cubic km \cite{design}.  The DOMs will be equally spaced on up to 80 strings, at depth from ~1.5 to ~2.5 km in the deep, clear Antarctic ice.  An array of surface stations, IceTop, enhance the ability to trigger on, or veto, down-going showers. Each IceTop station consisted of two clear ice tanks, each instrumented with 2 DOMs.  An IceTop station is located roughly 10 meters from each bore hole of the In-Ice array.

\begin{figure}[h]
\begin{center}
\includegraphics[width=18pc]{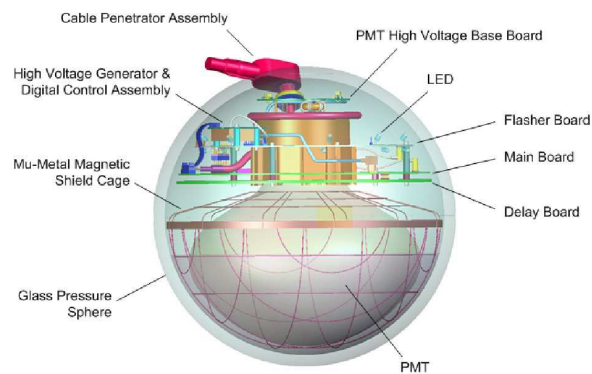}
\caption{\label{dom}A schematic view of an IceCube Digital Optical Module (DOM)}
\end{center} 
\end{figure}

IceCube is designed to detect Cherenkov radiation photons emitted by charged particles. The particles that are most likely to penetrate through the 1.5 km of ice on top of the detector are muons (from interaction of cosmic rays in the atmosphere), and neutrinos (atmospheric or from any other source). The rate of the muon background is about 6 orders of magnitude larger than that of atmospheric neutrinos, and therefore neutrino searches are performed using up-going particles that traverse the entire earth, using the matter of our planet as a muon screen. Based on measurements of the number of photons arriving at different DOMs and their arrival times, a track or a cascade can be reconstructed.
Each DOM is an autonomous data collecting and analyzing unit consisting of a 10" Hamamatsu PMT in a 12" 
pressure sphere (see figure \ref{dom}). A main board inside the DOM can digitize up to 300 
Mega Samples per Second (MSPS) for 400 
ns and 40 MSPS for $6.4 \mu s$.  A flasher board, populated with 12 Light Emitting Diodes (LEDs), produces 
pulses used for optical and timing calibration.
The 
DOMs can operate in a local coincidence mode, where a data recording will be triggered only if its neighbors were triggered within a certain time window. 

The main scientific goal of IceCube is to map the neutrino sky \cite{design}. IceCube will also look for high energy GZK neutrinos \cite{GZK}, study air showers, high energy atmospheric neutrinos \cite{atmospheric} and look for supernovas in a special data acquisition mode \cite{supernova}.  IceCube measurements can be used, to some extent, also for neutrino mass hierarchy and CP phase measurements \cite{winter}. There are models predicting certain neutrino flux enhancements, which could be measured by IceCube. These sources include, but are not limited to, dark matter, super-symmetry, magnetic monopoles, quantum gravity \cite{exotic}.

\begin{figure}[h]
\begin{minipage}{18pc}
\includegraphics[width=18pc]{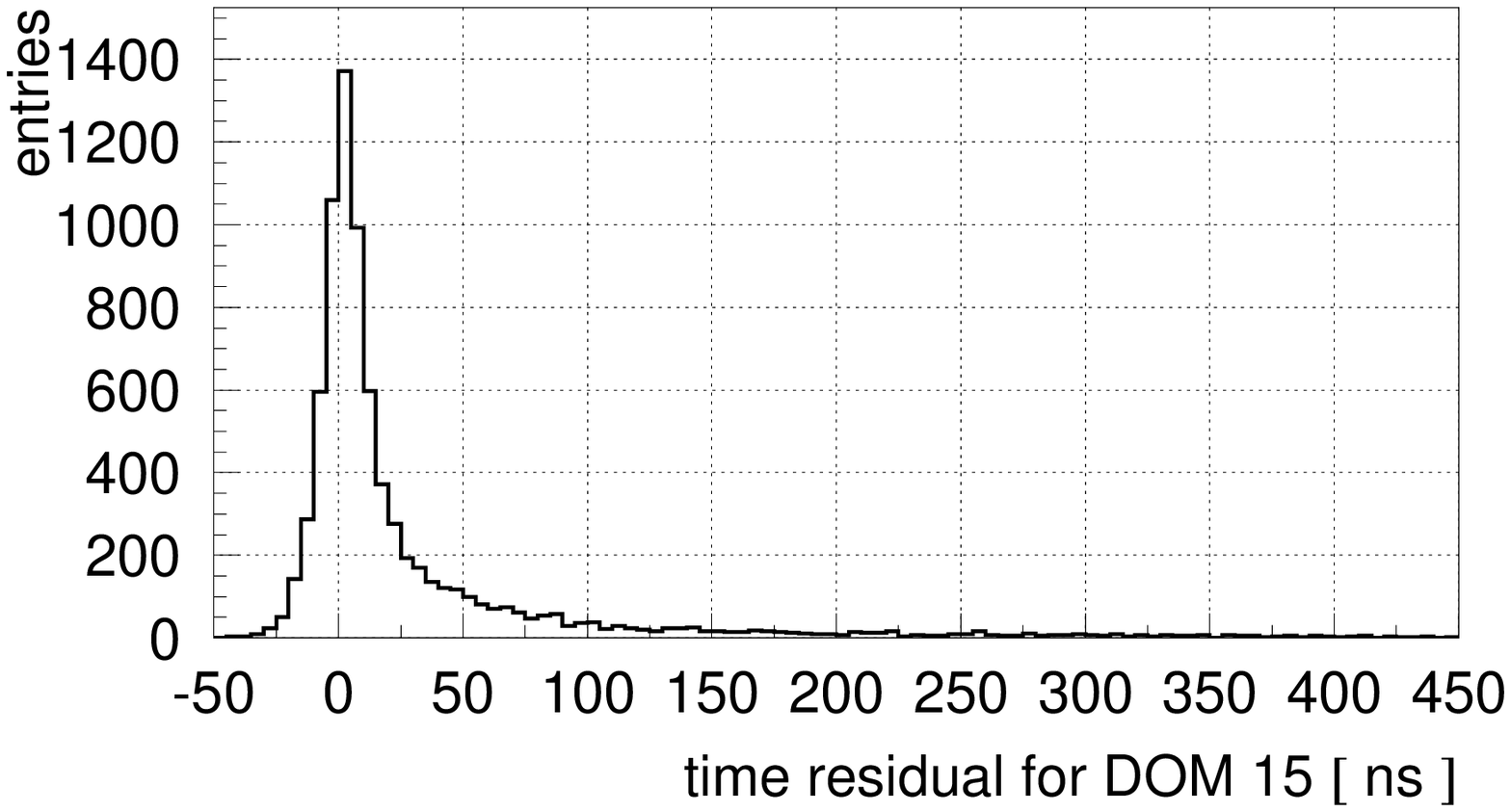}
\caption{\label{timeRes1}
Distribution of time residuals of
photons arriving at a DOM from nearby tracks reconstructed with the rest of the
string. Photons arriving from the track directly with no scattering will have time residual = 0}
\end{minipage}\hspace{2pc}%
\begin{minipage}{18pc}
\includegraphics[width=18pc]{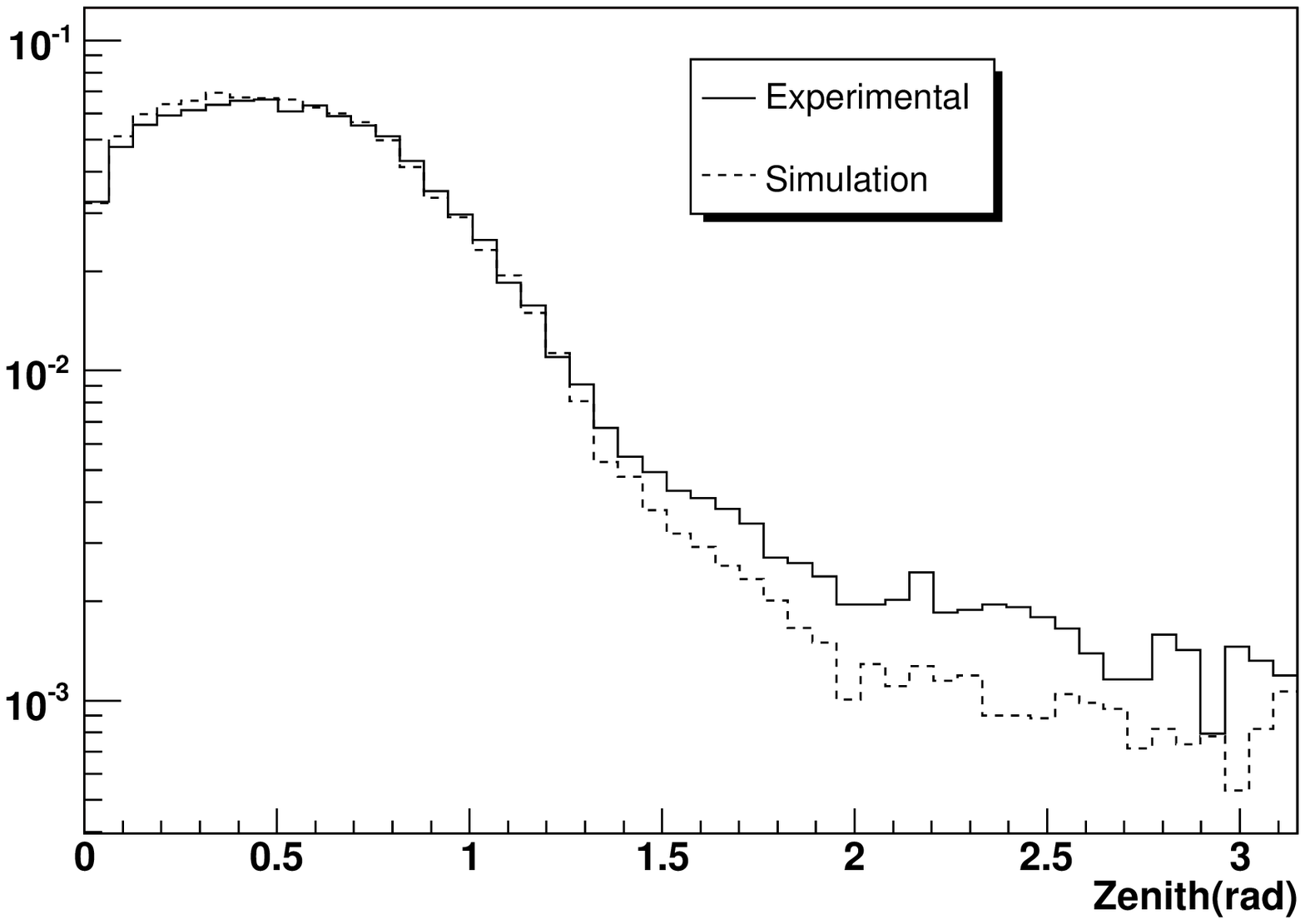}
\caption{\label{mcComp2}Distribution of reconstructed zenith angle for muon events }
\end{minipage} 
\end{figure}

\section{Current Status and Verification}
In the winter of 2004-2005 a single In-Ice string and 4 IceTop stations 
were deployed. At the end of the 2006 Austral summer IceCube consists of 9 In-Ice strings and 16 IceTop stations. A set of measurements were performed to confirm the design goal of the detector and check its performance \cite{firstyear}. In order to reconstruct and time tracks over the entire array, a timing resolution of a few nanoseconds is needed. The time resolution of each detector unit was estimated in two independent ways. In the first the flasher board was used. A LED on a DOM was flashed and adjacent DOMs triggered on it.  The time delay between the flashing and the triggering was measured multiple times. This procedure was repeated for all DOMs, and the maximum time RMS resolution was found to be less than 2ns.  A different way to estimate the time resolution is by reconstructing down-going muon tracks excluding one DOM, and calculating the time difference between the measured hit time and the expected hit time. Figure \ref{timeRes1} shows the distribution of those time residuals for a single DOM using multiple events. The process is repeated for all DOMs. The resolution was found to be less than 3 ns, after correcting for ice properties.  

The distributions of different reconstruction parameters were compared to predicted rates and simulated Monte Carlo events. For both down-going atmospheric muon candidate events, and up-going neutrino candidates events the agreement between Monte Carlo and data was good, and in agreement with previous predictions \cite{sensetivity}. A comparison of the zenith angle distributions is shown in figure \ref{mcComp2}.
In order to estimate signal and background behavior and check detector performance, a reliable neutrino interaction and detector simulation is needed. The IceCube simulation software is currently under active development. 

\begin{figure}[h]
\begin{minipage}{18pc}
\includegraphics[width=18pc]{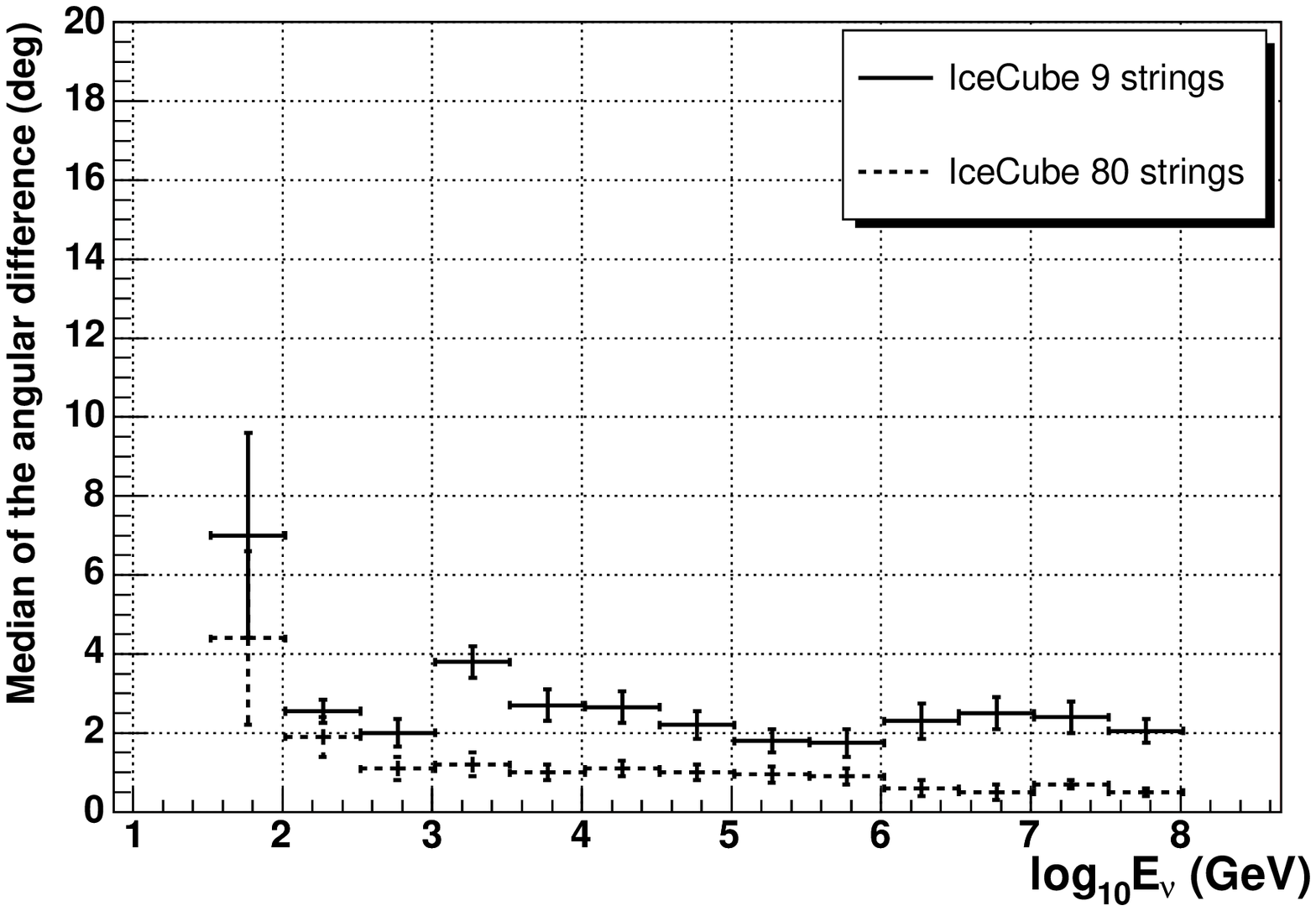}
\caption{\label{EnergyRes}The angular resolution for 9 In-Ice strings and for the full IceCube array as function of event energy (Preliminary). Shown here the differences between the true and reconstructed muon track.}
\end{minipage}\hspace{2pc}%
\begin{minipage}{18pc}
\includegraphics[width=18pc]{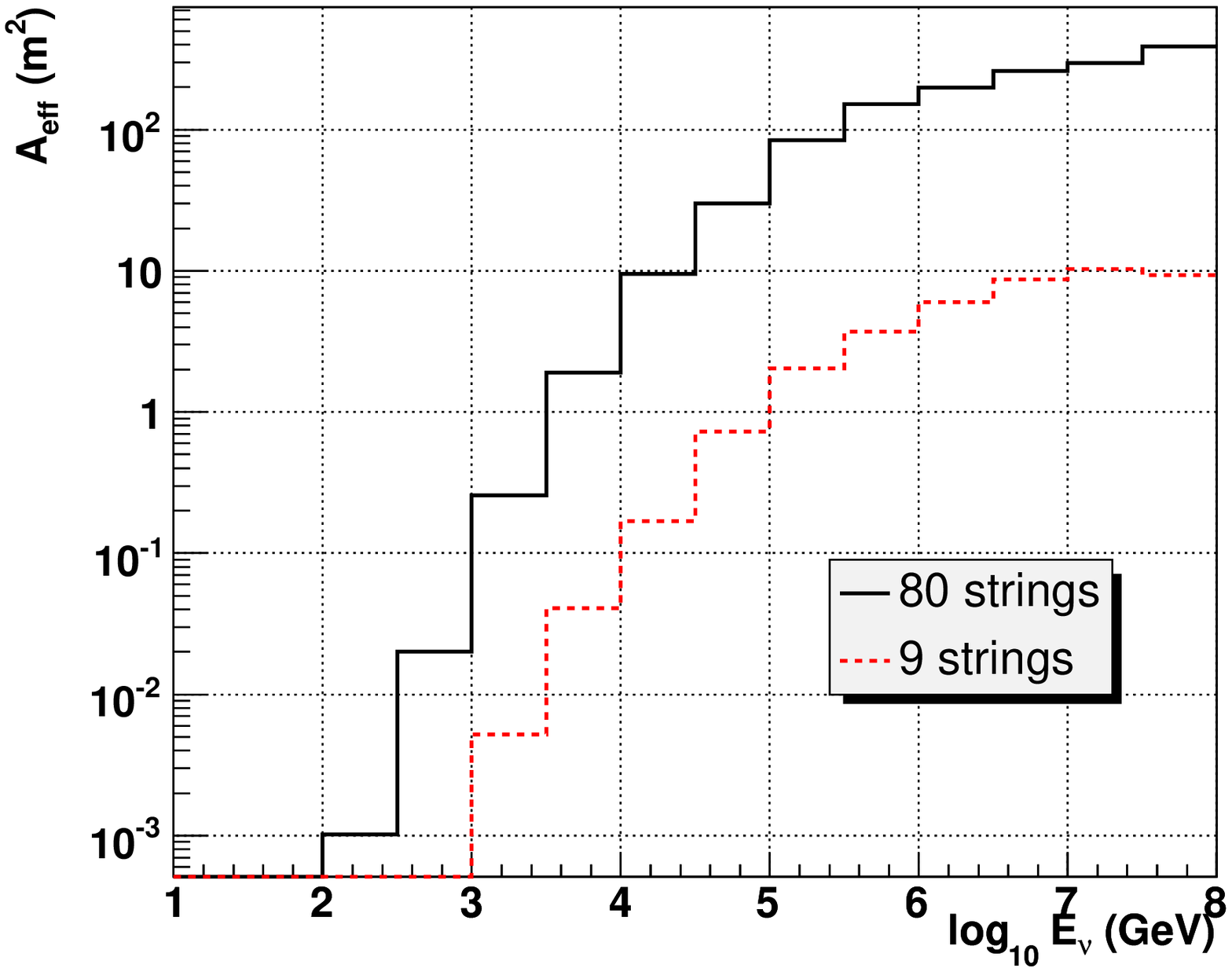}
\caption{\label{area}The effective area calculated for 9 and 80 strings detector, for reconstructed muon tracks as function of the neutrino energy (Preliminary)   }
\end{minipage} 
\end{figure}

\section{Future Sensitivity}

Using simulation of the 9 and 80 string detector configurations, effective areas, angular resolutions and event rates were estimated.  
Likelihood reconstructions run on simulation are used to estimate the angular resolution of the detector, shown as a function of energy in figure \ref{EnergyRes}.  
Resolutions can also be characterized by energy spectral indices. For 
example, the angular resolution for reconstructed atmospheric muon 
tracks will be about $2.2^\circ$ ($2.7^\circ$)  using 80 (9) strings, 
and $0.8^\circ$ ($1.3^\circ$) for an $E^{-2}$ spectrum. 
The angular resolution results quoted are expected to improve when the DOM waveform information will be fully used.
In figure \ref{area} the estimated effective area for neutrino detection as function of energy is shown.

\newpage 

\setcounter{section}{0}
\title{Multi-year search for a diffuse flux of muon neutrinos with AMANDA-II}

\author{J. Hodges for the IceCube Collaboration}

\address{Department of Physics, University of Wisconsin, Madison, Wisconsin, USA}

\ead{hodges@icecube.wisc.edu}

\begin{abstract}
A search for TeV to PeV muon neutrinos from unresolved sources was
performed on AMANDA-II data collected between 2000 to 2003. The diffuse
analysis sought to identify an extraterrestrial neutrino signal on top of
the atmospheric muon and neutrino backgrounds. An upper limit of
\mbox{$E^{2}\Phi_{90\% C.L.} < 8.8 \times 10^{-8}$ GeV cm$^{-2}$ s$^{-1}$
sr$^{-1}$} was placed on the diffuse flux of muon neutrinos with a dN/dE
$\sim$ E$^{-2}$ spectrum for the energy range 15.8 TeV to 2.5 PeV. Limits
were also placed on prompt and astrophysical neutrino models with other
energy spectra.
\end{abstract}

\section{Introduction}
Current theories on cosmic particle acceleration predict that neutrinos and
gamma rays are among the by-products of pp and p$\gamma$ interactions in
sources such as AGN (active galactic nuclei) or GRBs (gamma ray
bursts). Many extraterrestrial TeV gamma ray sources have already been
identified by other experiments, but the missing link is the detection of
an extraterrestrial neutrino flux. This search was optimized to look for
extraterrestrial neutrinos with a \mbox{dN/dE $\sim$ E$^{-2}$} spectrum,
the most general prediction from first order Fermi acceleration models.

A diffuse search for neutrinos does not use specific time or location
information. Instead, it looks for an excess of events over a large sky
region over a long period of time. If the neutrino flux from an individual
source is too small to be detected by current means, it is possible that
many similar sources, isotropically distributed throughout the Universe,
would combine to make a detectable signal. An excess of events over the
expected atmospheric neutrino background would be indicative of an
extraterrestrial neutrino flux.

\section{Search Methods}

Data for this analysis were collected by AMANDA-II between 2000 to
2003. This period covered 807 days of stable detector livetime. During this
period, 5.2 $\times$ 10$^{9}$ events triggered AMANDA-II.

\subsection{Backgrounds for the diffuse analysis}

Several types of events that can trigger the detector were
simulated. Atmospheric muons and neutrinos created when cosmic rays
interact with the Earth's atmosphere are the main background to
extraterrestrial neutrino-induced events. Atmospheric and extraterrestrial
neutrinos can travel from the far side of the Earth, interact in the ice or
rock near the detector, and induce an upward-moving muon that can be
detected. Atmospheric muons, on the other hand, do not have enough energy
to travel a long distance through the earth, and hence they can only
trigger the detector if they travel downward from the polar surface into
the ice.

\begin{figure}[h]

\includegraphics*[width=0.47\textwidth]{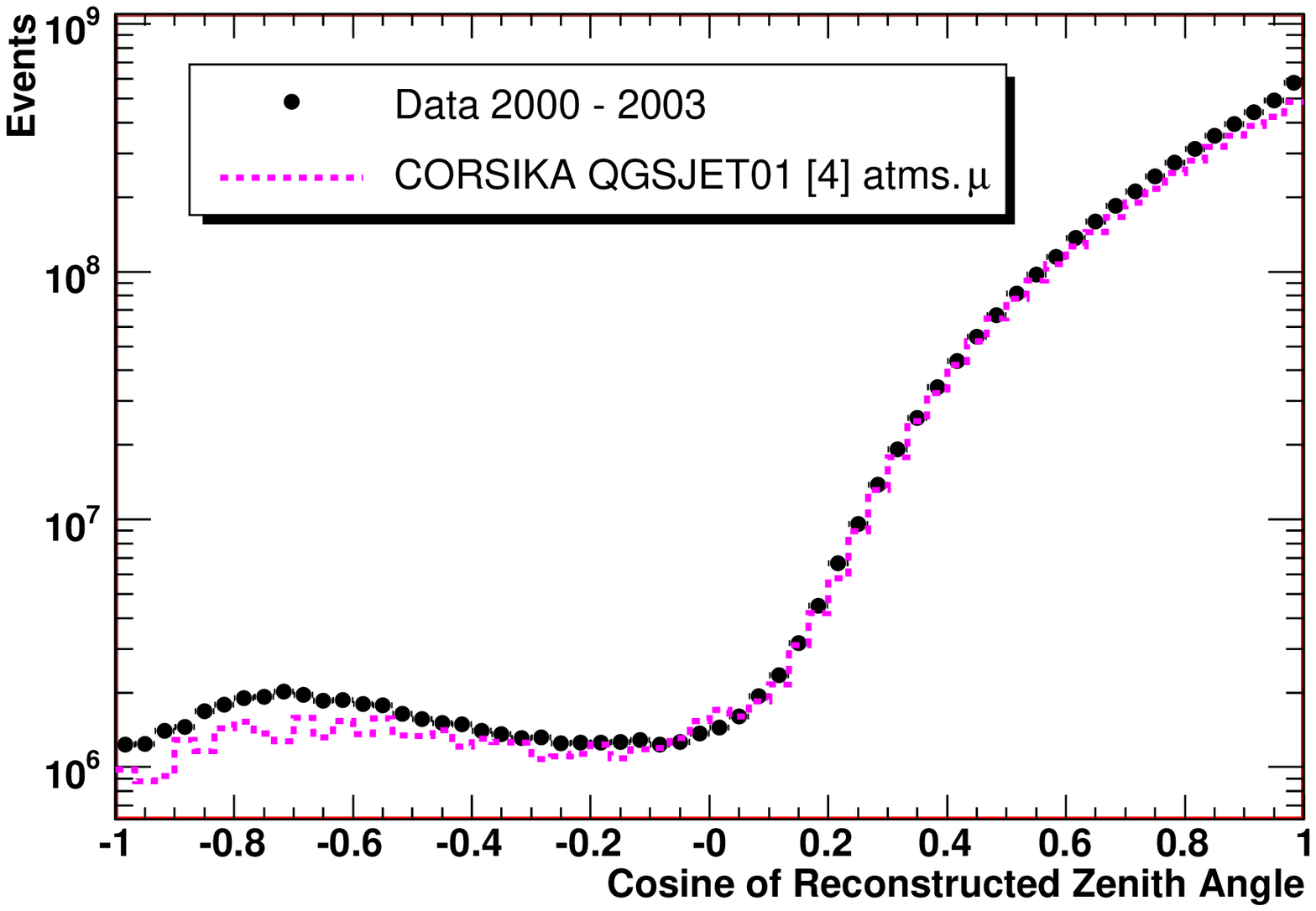}
\includegraphics*[width=0.47\textwidth]{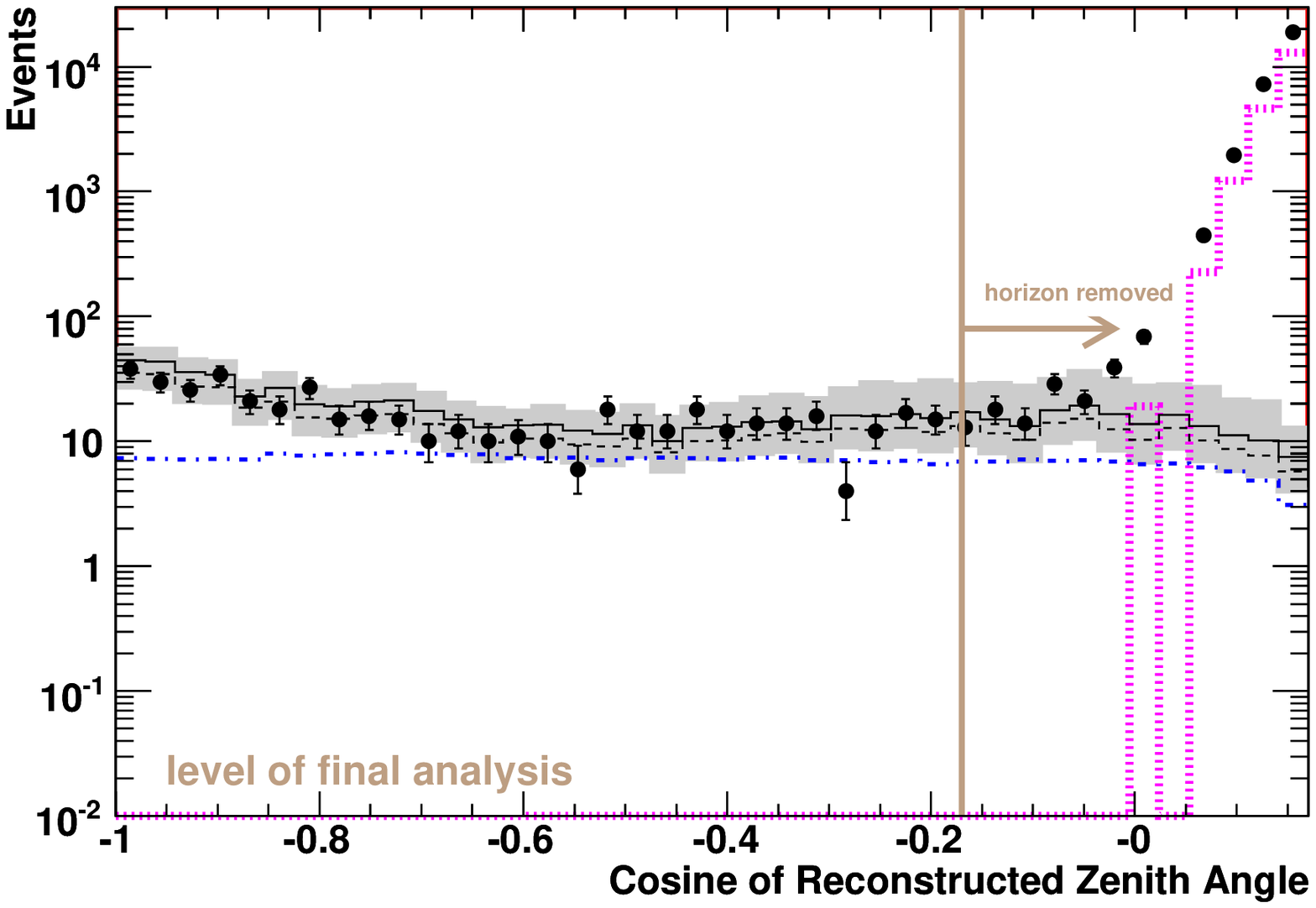}

\mbox{
\begin{minipage}[!t]{10cm}\caption{\label{coszenplots} The cosine of the
zenith angle is plotted for all events that triggered the detector
(left). Events at \mbox{cos(zenith) = -1} are traveling straight up through
the detector from the Northern Hemisphere. On the right, event quality
requirements were used to select the best upgoing tracks. The normalization
of the atmospheric neutrino simulation was adjusted so that the number of
events hitting between 50 and 100 OMs was the same in the data and
simulation.}
\end{minipage}

\begin{minipage}[!t]{5cm}
\includegraphics*[width=0.99\textwidth]{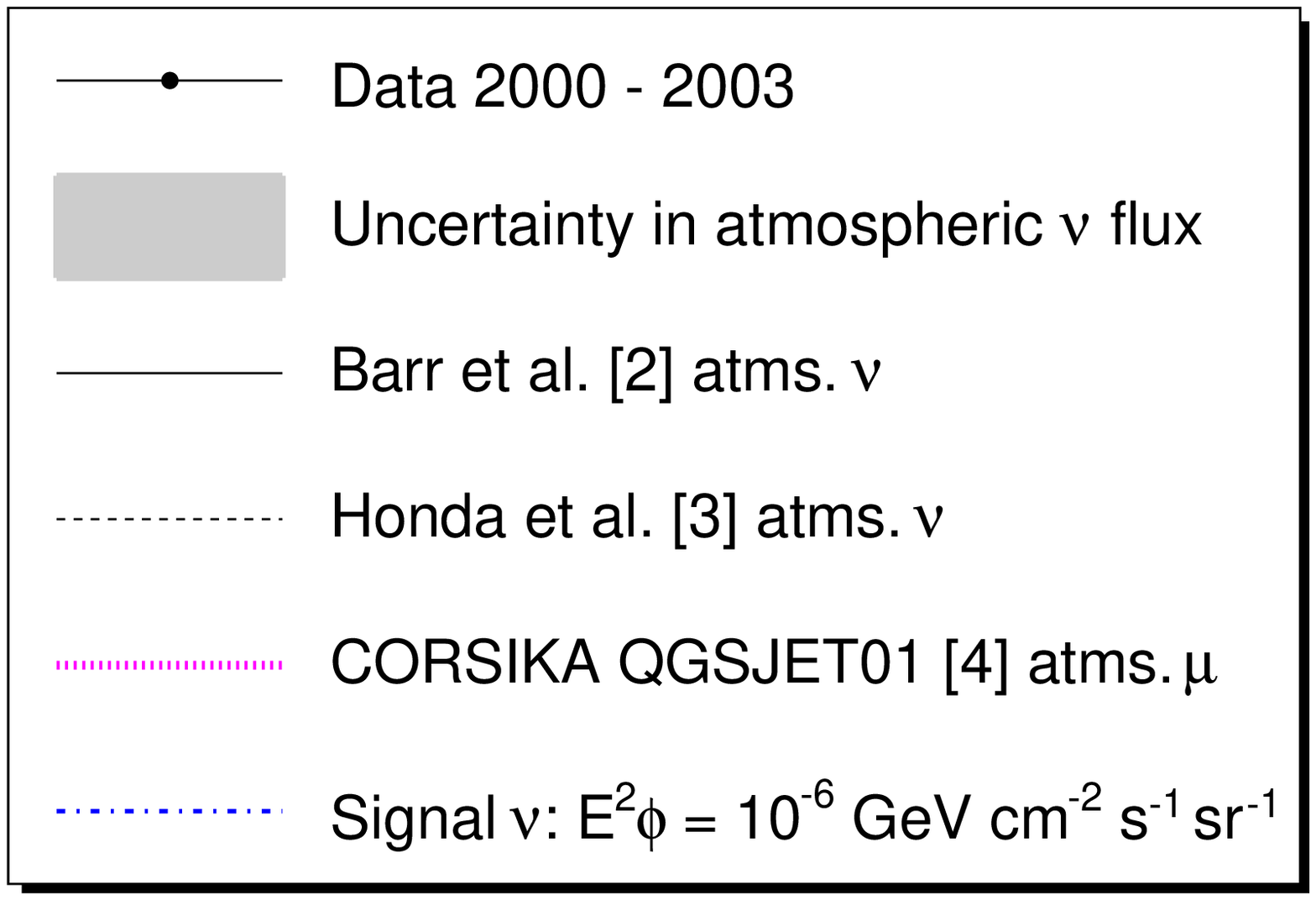}
\end{minipage}
}
\end{figure}

The first step in the analysis was to guess an arrival direction for every
event \cite{nim2004}, as shown on the left in Figure
\ref{coszenplots}. Using this directional information, all events that were
reconstructed in the downgoing direction were removed. The Earth was used
as a filter and the actual search for extraterrestrial neutrinos was only
performed on upgoing events.

Since the arrival direction of many downgoing atmospheric muons was
originally misreconstructed, event quality requirements were
introduced. Events were required to have long, smooth tracks of light that
had many Cherenkov photons arriving close to their expected arrival
times. This helped remove any misreconstructed downgoing events and helped
to assure a purely upgoing sample that can be seen on the right in Figure
\ref{coszenplots}.

\subsection{Separating atmospheric neutrinos from extraterrestrial neutrinos}

Atmospheric neutrinos from pions and kaons (\mbox{dN/dE $\sim$ E$^{-3.7}$})
have a softer energy spectrum than the proposed extraterrestrial neutrino
signal (\mbox{dN/dE $\sim$ E$^{-2}$}). As a result, these two event classes
can be separated best by their energy. At high energy, the extraterrestrial
neutrino flux would dominate over the atmospheric neutrinos.

\begin{figure}[ht]
\centering
\mbox {  

\begin{minipage}[!b]{7.6cm}
\includegraphics[clip,width=0.99\textwidth]{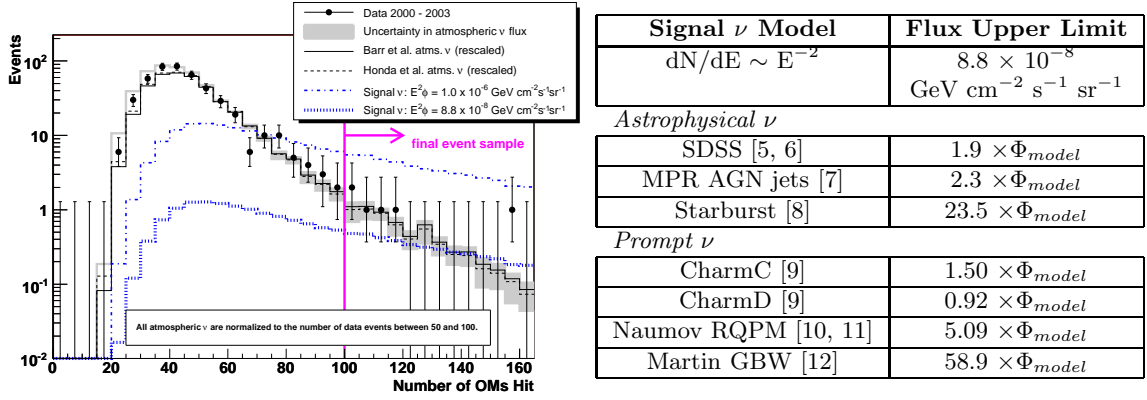}
\end{minipage}

\hfill

\begin{minipage}[!b]{6.175cm}
\footnotesize
\begin{tabular}{|c|c|}
\hline
\textbf{Signal $\nu$ Model} & \textbf{Flux Upper Limit} \\
\hline
\hline
dN/dE $\sim$ E$^{-2}$ & 8.8 $\times$ 10$^{-8}$ \\
 & GeV cm$^{-2}$ s$^{-1}$ sr$^{-1}$\\
\hline
\multicolumn{2}{l} {\emph{Astrophysical $\nu$}} \\
\hline
SDSS \cite{sdss,sdss_revision}& 1.9 $\times \Phi_{model}$\\
\hline
MPR AGN jets \cite{mpr}& 2.3 $\times \Phi_{model}$\\
\hline
Starburst \cite{loeb_waxman_starburst} & 23.5 $\times \Phi_{model}$\\
\hline
\multicolumn{2}{l} {\emph{Prompt $\nu$}} \\
\hline
CharmC \cite{zhv_charm} & 1.50 $\times \Phi_{model}$\\
\hline
CharmD \cite{zhv_charm} & 0.92 $\times \Phi_{model}$\\
\hline
Naumov RQPM \cite{naumov_rqpm_a, naumov_rqpm_b}& 5.09 $\times \Phi_{model}$\\
\hline
Martin GBW \cite{martin_gbw} & 58.9 $\times \Phi_{model}$\\
\hline
\end{tabular}
\end{minipage}
}
\caption{\label{nchplot}(Left) The number of OMs triggered during each event is an
energy-correlated observable. Events triggering at least 100 OMs appeared
in the final data set, and the number of observed data events was compared
to the simulated atmospheric neutrino background. (Right) Upper limits were
also determined for flux models ($\Phi_{model}$) with different energy
spectra. }

\end{figure}

Since the energy of an event is not directly observable, the number of
optical modules (OMs) hit during an event was used as an energy-correlated
parameter. Optimization studies performed on the simulation indicated that
the best signal-to-background region would be obtained by using events with
at least 100 OMs triggered. The number of data events seen in this high
energy window was compared to the predicted atmospheric neutrino
background, shown in Figure \ref{nchplot}.

\section{Systematic uncertainties}

An extensive systematic uncertainty analysis was performed to include
uncertainties in the neutrino flux models and detector performance. Two
different atmospheric neutrino models were used, Barr
\textit{et al.} \cite{bartol2004} and Honda \textit{et al}
\cite{honda2004}. Uncertainties in the cosmic ray flux and the hadronic
interaction model were also considered. All of the atmospheric neutrino
simulation was scaled so that the number of simulation events matched the
number of data events in the region 50 \textless number of OMs hit
\textless 100.

To assess detector and simulation performance, an inverted analysis was
performed in which the highest quality downgoing events were
studied. Downgoing events that were previously eliminated (0$^{o}$ \textless
zenith angle \textless 80$^{o}$) were reintroduced. With very high statistics
available from these downgoing events, the characteristics of high energy
events were studied without having to reveal the high energy upgoing data
events.

\section{Results}

Six data events were observed on an average predicted atmospheric neutrino
background of 6.1 events. Since no excess of events was seen indicating an
extraterrestrial signal, an upper limit was set for a dN/dE $\sim$ E$^{-2}$
flux between 15.8 TeV to 2.5 PeV (the energy region covered by 90\% of the
simulated signal). The upper limit on the diffuse flux of muon neutrinos
from AMANDA-II data from 2000 to 2003 is \mbox{$E^{2}\Phi_{90\% C.L.} < 8.8
\times 10^{-8}$ GeV cm$^{-2}$ s$^{-1}$ sr$^{-1}$}.

Signal models with other energy spectra were also tested with this
data. Due to the different nature of their energy spectra, the requirement
of how many OMs were triggered during an event was reoptimized. The upper
limit on each of the models appears in the table above.

\newpage
\setcounter{section}{0}
\title{Searches for Neutrinos from Gamma Ray Bursts with AMANDA-II and IceCube}

\author{B. Hughey for the IceCube Collaboration }

\address{Department of Physics, University of Wisconsin, Madison, Wisconsin, USA}

\ead{hughey@icecube.wisc.edu}

\begin{abstract}
The hadronic fireball model predicts a neutrino flux in the TeV to several PeV range simultaneous with the prompt photon emission of GRBs.  The discovery of  high energy neutrinos in coincidence with a gamma ray burst would help confirm the role of GRBs as accelerators of high energy cosmic rays.  We summarize the methods employed by the AMANDA experiment in the search for neutrinos from GRBs and present results from several analyses.  
\end{abstract}

\section{Neutrinos From GRBs}

Gamma Ray Bursts (GRBs) are one of the most plausible sources
of ultra-high energy cosmic rays \cite{wb,crgrb2}. Detection of high energy neutrinos from a burst would provide corroborating evidence for the production of ultra-high energy cosmic rays inside GRBs.

It is believed that gamma rays produced by GRBs originate from electrons
accelerated in internal shock waves associated with relativistic jets
(with Lorenz boost $\Gamma$$\sim$300). These gamma rays have energies in
the range from 10 keV  to greater than 10 MeV. The gamma ray spectrum can be described as a broken power law, with a softer spectrum above a
break energy which is typically 0.25-1 MeV. Gamma ray bursts can
last anywhere from a few milliseconds up to a few hundred seconds.  The distribution of durations is usually considered to be composed of two separate classes, with short bursts lasting less than 2 seconds and long bursts lasting more than 2 seconds \cite{batse}. Gamma ray bursts are reviewed in \cite{grbreview} and \cite{piranreview}.

If protons and/or nuclei are also accelerated in the jets, then
high energy neutrinos ($\sim 10^{14}$~eV) are produced \cite{wb} via the
process:

\begin{equation}
\label{eqn:pgamma}
p+\gamma \rightarrow \Delta^+ \rightarrow \pi^+[+n] \rightarrow \nu_\mu +
\mu^+ \rightarrow \nu_\mu + e^+ + \bar{\nu}_{\mu} + \nu_{e}.
\end{equation}

The neutrino flavor ratio
$\nu_{e}$:$\nu_{\mu}$:$\nu_{\tau}$ is thus 1:2:0 at source. Taking into account neutrino oscillations, the flavor ratio observed at Earth is 1:1:1
\cite{athar}. However, Kashti and Waxman \cite{kashti} point out that at energies greater than
$\sim$~1~PeV, the $\mu^{+}$ in Equation~(\ref{eqn:pgamma}) loses energy through synchrotron radiation before 
decaying. This energy loss changes the source
neutrino flavor ratio at high energies from 1:2:0 to 0:1:0, leading to a ratio at Earth of 1:1.8:1.8. 

Neutrino production is predicted to be simultaneous with gamma
ray production.  AMANDA GRB analyses use the  Waxman-Bahcall \cite{wb} broken power law neutrino spectrum as a reference hypothesis (see Fig. 1). However, other models of prompt neutrino emission have also been tested.  These include the paramaterization of  Murase and Nagataki \cite{murase}, who arrive at a similar spectrum to Waxman-Bahcall under different assumptions, as well as the supranova scenario (now disfavored due to evidence from the Swift satellite) which assumes GRB jet interactions with an external matter field created by a supernova preceding the burst by $\sim$1 week \cite{supranova}.  Predictions have also been made for precursor \cite{precursor} and afterglow \cite{afterglow} emission.

\section{The AMANDA Detector}

The Antarctic Muon and Neutrino Detector Array (AMANDA)  \cite{ama:nature-pub,nu2004:ama} is located at the South Pole.  From 1997 to 1999, AMANDA consisted of 302 optical modules on 10 strings and was referred to as AMANDA-B10.  The final configuration, AMANDA-II, was commisioned in the year 2000 and consists of a
total of 677 optical modules on 19 strings.  Each module contains a photomultiplier tube and
supporting hardware inside a pressurized glass sphere.  The
optical modules are used to indirectly detect neutrinos by measuring the
Cherenkov light from secondary charged particles produced in neutrino-nucleon
interactions.  

AMANDA uses two detection channels.  Muon tracks are produced through interactions of $\nu_\mu$, while cascades (particle showers) are produced from interactions of all three neutrino flavors.  The muon channel has a larger effective areas because of the longer range of muons compared to cascades. It also has better pointing resolution because muons produce linear tracks rather than spherical showers.  Separating neutrino signals from the dominant atmospheric muon background is accomplished by removing downgoing events, so muon analyses have $\sim 2\pi$ sr sky coverage.  Cascades are differentiated from downgoing muons by their shape and therefore cascade analyses have full (4$\pi$ sr) sky coverage. Cascade events also have better energy resolution than muon tracks, since the energy of all particles produced in the shower is accounted for.

\section{AMANDA GRB Analyses}

In the majority of GRB analyses, searches are done in coincidence with $\gamma$-ray detections by satellites.  Because these analyses only search for a neutrino signal during  the time and (in the case of muon channel searches) in the location of measured bursts, there is almost no on-source background in these analyses.  The period of time actually examined for a neutrino signal for each burst is equal to the measured duration of prompt gamma-ray emission, plus the uncertainty in this measurement, plus an additional second on each side of the on-time window.  Background was measured for a period of one hour both before and after each burst, with the ten minute period immediately surrounding the burst remaining unexamined to avoid the possibility of contaminating the background with neutrino signal.   In the muon channel, searches for prompt emission have been conducted for 312 bursts measured by the BATSE detector (aboard the CGRO satellite) and 95 bursts analyzed by the IPN3 satellite.  Additionally, a search for precursor emission was conducted using 60 bursts from the 2001-2003 data sets \cite{ama:grb-muon}.  Using the cascade channel, 73 bursts identified by the BATSE detector in the year 2000 have been studied \cite{ama:grb-casc}.   No events have been observed in coincidence with any bursts studied so far, which is consistent with the expected background.

The \emph{rolling} analysis provides a useful complement to these triggered searches.  This method does not use satellite triggers, but scans an entire multi-year data sample for a statistical excess of events within one of two pre-set time windows (to account for both long and short burst classes).  This allows this analysis to search for GRBs and other transients not identified by satellites.  The rolling search has been conducted for the years 2001-2003 (after the BATSE detector was ceased operations and before the Swift satellite launched) using the cascade channel. Due to the larger amount of data analyzed relative to the triggered analyses, more stringent cuts on the data are required.  Thus, background rejection was accomplished with a six cut-variable Support Vector Machine, optimized for the best chance for signal discovery.  As in the case of the triggered searches, no evidence of astrophysical neutrinos has been found with this analysis method.  The maximum number of observed events and the numbers of observed windows with multiple (2 or 3) events is consistent with the predicted background \cite{ama:grb-casc}.

Although the Waxman-Bahcall neutrino spectrum functions as a reference for GRB analyses, it has been demonstrated that neutrino spectra from individual bursts can vary significantly from this ``standard'' spectrum \cite{becker,guetta}.  Current AMANDA analyses are using more sophisticated methods to predict the spectrum and neutrino rates for individual bursts rather than assuming averaged parameters.  The particularly close and bright burst GRB030329 was the first burst to be given this individualized treatment \cite{icrc05:grb}.  Bursts detected by Swift, many of which have redshifts directly measured from afterglow data, will be especially conducive to this method.

IceCube, the successor to AMANDA, is currently under construction, with the final detector scheduled be completed by 2011. Preliminary studies indicate that a triggered search using 300-500 bursts with the full IceCube array would suffice to either set limits at levels lower than the predictions by Waxman-Bahcall or find evidence of the existence of neutrinos in coincidence with GRBs with better than 5$\sigma$ confidence.

\begin{figure}
\begin{center}
\mbox{
\includegraphics*[width=.43\textwidth]{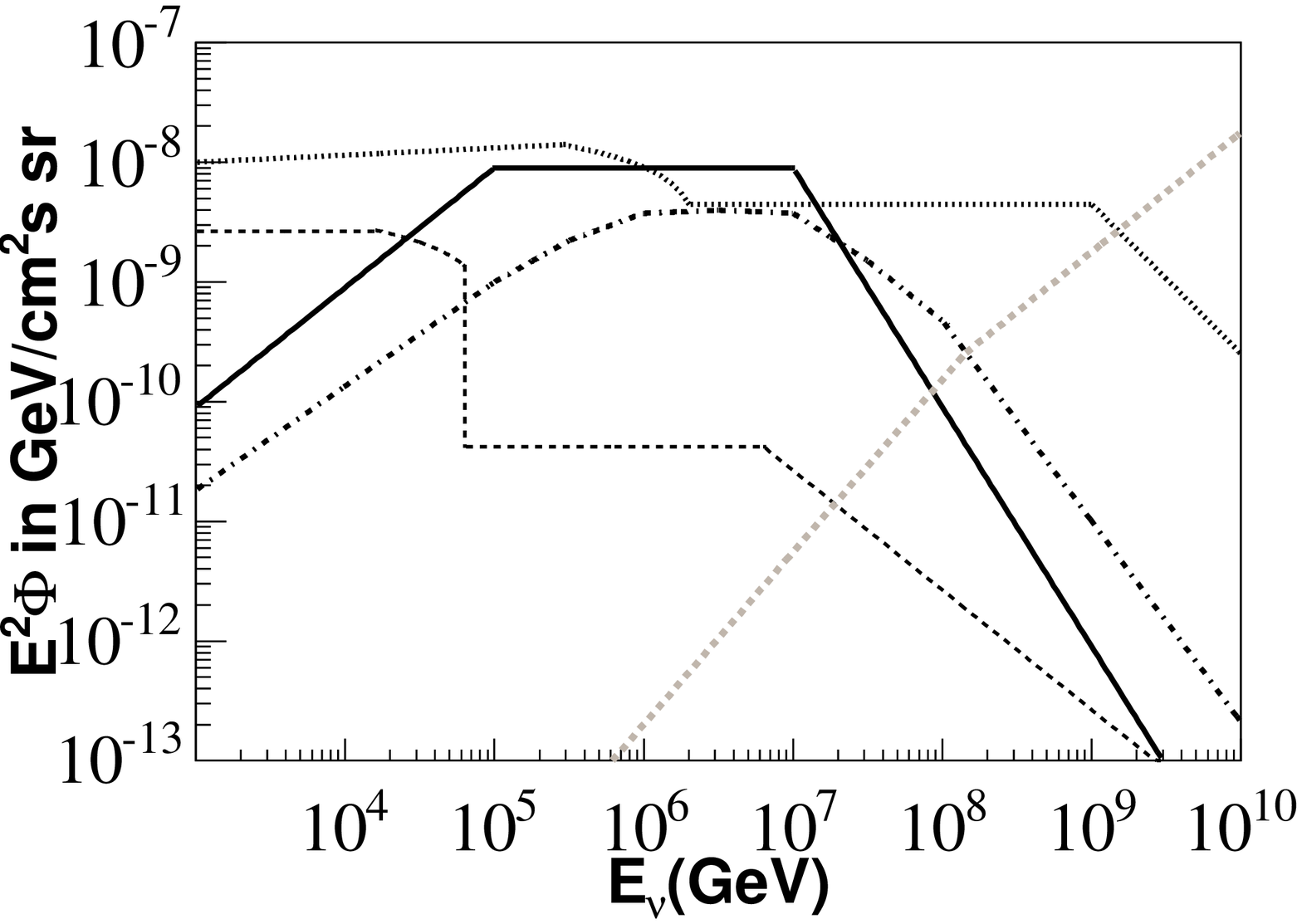}}
\mbox{
\includegraphics*[width=.43\textwidth]{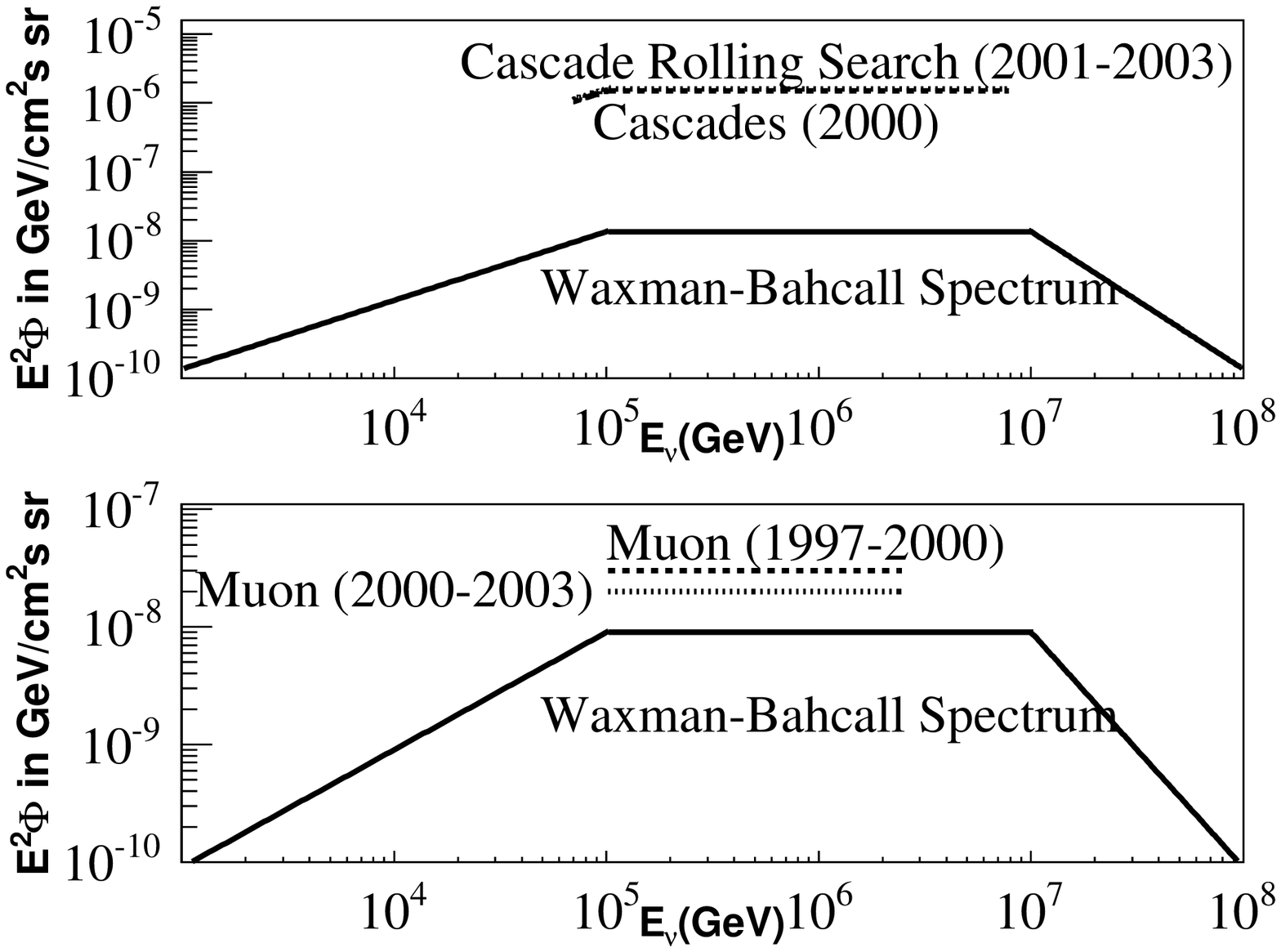}}
\caption{\label{fig:1}Left: Models of neutrino emission from GRBs: Solid: Waxman-Bahcall; Dotted: Supranova; Dash-Dotted: Murase-Nagataki; Light Dashed: Afterglow; Dark Dashed: Precursor Emission. Right: Experimental limits relative to Waxman Bahcall spectrum for cascade(top) and muon(bottom) channels, with displayed ranges containing 90\% of simulated signal events.} 
\end{center}
\end{figure}

\newpage

 \begin {paragraph}
{\bf Acknowledgments}
 We acknowledge the support from the following agencies: National Science Foundation-Office of Polar Program, National Science Foundation-Physics Division, University of Wisconsin Alumni Research Foundation, Department of Energy, and National Energy Research Scientific Computing Center (supported by the Office of Energy Research of the Department of Energy), the NSF-supported TeraGrid system at the San Diego Supercomputer Center (SDSC), and the National Center for Supercomputing Applications (NCSA); Swedish Research Council, Swedish Polar Research Secretariat, and Knut and Alice Wallenberg Foundation, Sweden; German Ministry for Education and Research, Deutsche Forschungsgemeinschaft (DFG), Germany; Fund for Scientific Research (FNRS-FWO), Flanders Institute to encourage scientific and technological research in industry (IWT), Belgian Federal Office for Scientific, Technical and Cultural affairs (OSTC); the Netherlands Organisation for Scientific Research (NWO); M. Ribordy acknowledges the support of the SNF (Switzerland); J. D. Zornoza acknowledges the Marie Curie OIF Program (contract 007921). 
 \end {paragraph}


\section*{References}
\begin{thebibliography}{9}
\bibitem{icepaper}
Ackermann M et.~al.
Optical properties of deep glacial ice at the South Pole
2006 \textit{J. Geophys. Res.} \textbf{111} D13203
\bibitem{amanda:nature} 
Andres E et.~al. 
Observation of high-energy neutrinos using Cherenkov detectors embedded
deep in Antarctic ice
2001 \textit{Nature} \textbf{410} 441--3
\bibitem{amanda:prd}
Ahrens J et.~al.
Observation of high energy atmospheric neutrinos with the Antarctic muon
and neutrino detector array
2002 \textit{Phys.~Rev.~D} \textbf{66} 012005
\bibitem{ic3:sens}
Ahrens J et.~al.
Sensitivity of the IceCube detector to astrophysical sources of high
energy muon neutrinos.
2004 \textit{Astropart. Phys.} \textbf{20} 507--32
\bibitem{spase-amanda}
Ahrens J et.~al.
Calibration and survey of AMANDA with the SPASE detectors
2004 \textit{Nucl. Instr. Meth.} \textbf{A522} 347--59
\bibitem{ic3:tev-icetop}
Bai X and Gaisser T
Air showers in a three dimensional array: recent data from IceCube/IceTop
2006 \textit{this proceedings}
\bibitem{ic3:performance}
Achterberg A et.~al.
First year performance of the IceCube neutrino telescope
2006 \textit{Astropart. Phys.} \textbf{26} 155--73
\bibitem{ic3:1st-res}
Dumm J and Landsman H
IceCube - first results
2006 \textit{this proceedings}
\bibitem{ic3:prod-test}
Hanson K and Tarasova O
Design and production of the IceCube digital optical module
2006 \textit{Proceedings of the 4th International Conference on New
Developments in Photodetection} published in 
\textit{Nucl. Instr. Meth.} \textbf{A567} 214--7
\end{thebibliography}

\section*{References}

\begin{thebibliography}{9}
\bibitem{stecker96} Stecker F W and Salamon M H 1996 {\it Space Science Reviews}
  {\bf 75} 341
\bibitem{nellen} Nellen L, Mannheim K and Biermann P L 1993 {\it Phys.~Rev.~D}
  {\bf 47} {5270}
\bibitem{mpr} Mannheim K, Protheroe R J and Rachen J P 2001 {\it Phys.~Rev.~D}
  {\bf 63} 23003
\bibitem{muecke} M\"ucke A et al. 2003 {\it Astrop.~Phys.} {\bf 18} 593
\bibitem{icrc05} IceCube Collaboration, contributions to ICRC 2005, Pune (India), {\it astro-ph/0509330}
\bibitem{andreas_stacking} Ackermann et al. 2006, ``On the selection of...'',
  accepted for publication in {\it Astrop.~Phys.}
\bibitem{point_source_5yr} Achterberg et al. 2006, to be submitted to {\it Phys.~Rev.~D}
\bibitem{jess_these_proc} Hodges J for the IceCube Collaboration 2006, these proceedings
\bibitem{diffuse_part} Becker J K, Rhode W, Biermann P L M\"unich K 2006 {\it astro-ph/0607427}
\bibitem{alvarez_meszaros} Alvarez-Mu{\~n}iz J and M{\'e}sz{\'a}ros P 2004
  {\it Phys.~Rev.~D} {\bf 70} 12, 123001
\bibitem{kael_these_proc} Hanson K for the IceCube collaboration 2006, these proceedings
\bibitem{francis_barca} Halzen F 2006, proc.~of {\it ``The multimessenger
  approach to high-energy $\gamma$-ray sources''},  Barcelona (Spain)
\end{thebibliography}

\section*{References}
\begin{thebibliography}{9}


\bibitem{thun1996}
M. Thunman, G. Ingelman and P. Gondolo, Astropart. Phys.{\bf 5}, 309 (1996) [arXiv:hep-ph/9505417]; L. Pasquali and M.H. Reno, Phys. Rev. D {\bf 59}, 093003 (1999) [arXiv:hep-ph/9811268]; J.F. Beacom and J. Candia, JCAP 0411, 009 (2004) [arXiv:hep-ph/0409046]; S.I. Dutta, M.H. Reno and I. Sarcevic, Phys. Rev. D {\bf 62}, 123001 (2000) [arXiv:hep-ph/0005310]; H. Athar, K.M. Cheung, G.L. Lin and J.J. Tseng, Astropart. Phys. {\bf 18}, 581 (2003) [arXiv:hep-ph/0112222]; A.D. Martin, M.G. Ryskin and A.M. Stasto, Acta Phys. Polon. B {\bf 34}, 3273 (2003) [arXiv:hep-ph/0302140].

\bibitem{kash2006}
Kashti and Waxman, astro-ph/0507599 and Barenboim and Quigg, hep-ph/0301220. 

\bibitem{astro2006}
IceCube Collaboration, Astropart. Phys. {\bf 26}, pp. 155-230 (2006).

\bibitem{lear1995}
J.G. Learned and S. Pakvasa, Astropart. Physics J. {\bf 3}, p. 267 (1995).   

\bibitem{deyo2006}
T. De Young, S. Razzaque and D.~.F.~Cowen, asXiv: astro-ph/0608486.

\end{thebibliography}

\section*{References}

\noindent
\begin{thebibliography}{99}

\bibitem{PS} Ackermann M et al 2006 Proceeding of The Multi-Messenger Approach to High Energy Gamma-ray Sources,
Barcelona, to be published in Astrophysics and Space Science Manuscript
\bibitem{stacking} Achtenberg A et al 2006  
 On the selection of AGN neutrino source candidates for a source stacking analysis with neutrino telescopes {\it Preprint} astro-ph/0609534
\bibitem{Begelman}
Begelman M C, Blandford R D, Rees M J 1984  \textit{Rev. Mod. Phys.} 56, 255 
\bibitem{SSC}
Jones, O'Dell,  Stein, 1974 \textit{ApJ} 188, 353 \\
Mastichiadis, Kirk J 1997 \textit{A\&A}, 320, 19 

\bibitem{K. Mannheim}
Mannheim K 1993 \textit{A\&A} 269, 67

\bibitem{Anita}
M\"ucke A., Protheroe R~J 2001 \textit{Astroparticle Physics} 15, 121-136\\
Aharonian F~A 2002 \textit{MNRAS} 215A, 332

\bibitem{ICRC} Ackermann M et al 2005,  Proceeding of 
29$^{th}$ International Cosmic Ray Conference (ICRC 2005) Pune (India) {\it Preprint} astro-ph/0509330, 24-27

\bibitem{EB} Bernardini E et al 2005, Proceeding of 7$^{th}$ Workshop of Towards a Network of Atmospheric Cherenkov
Detectors 2005, Palaiseau, France {\it Preprint} Astro-ph/0509396

\bibitem{ASM}
Levine A~M et al 1996 \textit{The Astrophysical Journal}, 469, L33-L36 

\bibitem{Martin}
Tluczykont M, Shayduk M, Kalekin O, Bernardini E, Long-term Gamma-Ray lightcurves and high-state
probabilities of Active Galactic Nuclei, these Proceedings

\bibitem{Teresa}
Bayer M, Larson K, Montaruli T, Steele D, 
Joint Multi-Wavelength Observations of Blazars with WIYN-VERITAS-IceCubeType, these Proceedings
\bibitem{Krawczynski501_2000}
Krawczynski H, Coppi P~S, Maccarone T, Aharonian F~A 2000 \textit{A\&A} 353, 97\\
Krawczynski H et al 2004 \textit{ApJ} 601, 151

\bibitem{scar98} Scargle J~D 1998 \textit{ApJ}  504, 405 
\bibitem{wolk05} Wolk S~J et al 2005 \textit{Astophys.J.Suppl}  160, 423 
\bibitem{internal} Resconi E, Gross A, Costamante L, Flaccomio E, Franco D, icecube/200608002

\end{thebibliography}

\medskip
\begin{thebibliography}{9}
%\bibitem{performance05} A.~Achterberg {\it et al.,} (IceCube Collaboration), 
%Astropart.~Phys. {\bf 26}, 155 (2006).
\bibitem{performance05} 
Achterberg A et al. (IceCube Collaboration) 
2006 {\it Astropart. Phys.} {\bf 26} 155

\bibitem{composition} 
Ahrens J et al.
2004 {\it Astropart.~Phys.} {\bf 21} 565

\bibitem{EASTOP-MACRO} 
Aglietta M et al. (EAS-TOP and MACRO Collaborations) 
2004 {\it Astropart.~Phys.} {\bf 20} 641; 
1994 {\it Phys.~Lett.}B {\bf 337} 376 

\bibitem{hillas06} 
Hillas A M 
{\it "Cosmic Rays: Recent Progress and some Current Questions"} 
arXiv:astro-ph/0607109 
\end{thebibliography}

\section*{References}
\begin{thebibliography}{99}

\bibitem{sgr1} Barat C., {\sl et al.}, {\it Astron. Astrophys.} {\bf 79},
L24-L25 (1979); Cline T., {\sl et al.} {\it IAU Circ.} No.7002 (1998)

 \bibitem{integral} Borkowski D. {\sl et al.}, {\it GCN Circ.} {\bf 2920},
 (2004); S. Mereghetti {\sl et al.}, {\it Astrophys. J.} {\bf 624}, L105
 (2005)

 \bibitem{swift-bat} Palmer D.M. {\sl et al.}, {\it GCN Circ.} {\bf 2925}, (2004)

 \bibitem{rhessi} Hurley K. {\sl et al.}, {\it Nature} {\bf 434}, 1098, (2005)

\bibitem{Kouveliotou} Kouveliotou C., Duncan R.C., Thompson C., {\it Sci. Am.} {\bf 288N2} (2003)

 \bibitem{Palmer}
 D.~M.~Palmer {\sl et al.}, Nature {\bf 434}, 1107 (2005).

\bibitem{ourtheory}
F.~Halzen, H.~Landsman and T.~Montaruli, 
eprint: astro-ph/0503348 and updated paper in preparation.

\bibitem{Gelfand} J.D.~Gelfand {\sl et al.},
Astrophys. J. {\bf 634}, 89 (2005).

\bibitem{Ioka} K. Ioka {\sl et al.},
Astrophys. J. {\bf 633}, 1013 (2005).

\bibitem{Andres} E. Andr\'{e}s {\sl et al.}, Astrop. Phys. {\bf 13},
 1 (2000).

\bibitem{geotail}
 T. Terasawa {\sl et al.},
 eprint: astro-ph/0502315, subm. to Nature.

 \bibitem{private} 
 D. Gotz, private communication.

\bibitem{cluster4} 
E. Pian and S. Schwartz, private communication.

\bibitem{hill}
 G. C. Hill, J. Hodges, B. Hughey, A. Karle and M. Stamatikos, Proc. of PHYSTAT 2005, Oxford. 
 G. C. Hill and K. Rawlins, Astrop. Phys., {\bf 19}, 393, (2003).

 \bibitem{corsika} D. Heck {\sl et al.}, FZKA-6019 (1998).

 \bibitem{anis}
  A. Gazizov and M.P. Kowalski, Comput. Phys. Commun. {\bf 172} 203 (2005). 

\end{thebibliography}

\section*{References}

\noindent

\begin{thebibliography}{99}
\bibitem{design} IceCube Project Preliminary Design Document, Ahrens et al. (IceCube Collaboration), http://icecube.wisc.edu.
\bibitem{GZK} R.Engel, D.Seckel and T.Stanev, Phys.Rev. D64 093010 (2001)
\bibitem{atmospheric} M.C. Gonzalez-Garcia, F.Halzen and M.Maltoni, Phys.Rev. D71, 092010 (2005).
\bibitem{winter} W. Winter, Phys.Rev. D74, 033015 (2006)
 \bibitem{exotic} L. Anchiordoqui et al., Phys. Rev. D72, 065019 (2005)
I.F. Albuquerque, G. Burdman and Z. Chako, Phys. Rev. Lett. 92, 221802 (2004)  
I.F. Albuquerque, J. Lamoreaux and G. Smoot, Phys, Rev. D66, 1215006 (2002)
\bibitem{firstyear} A.~Achterberg et al. (IceCube collaboration),  Astropar. Phys. 26,  155 (2006)
\bibitem{supernova} J.Ahrens et al. (AMANDA Collaboration), Astropart. Phys. 16  , 345 (2002)
\bibitem{sensetivity} J. Ahrens et al. (IceCube Collaboration),Astropart. Phys. 20 , 507 (2004)
%\bibitem{Bartol} V.Agrawal et al., Phys. Rev. D70 , 023006 (2004)
%\bibitem{diffuse} See ``Multi-year search for a diffuse flux of muon neutrinos with AMANDA-II'',  in these proceeding
\end{thebibliography}

\section*{References}

\begin{thebibliography}{99}

\bibitem{nim2004}
    Ahrens J \emph{et al} 2004 \textit{Nucl. Instr. Meth. A} \textbf{524} 169

\bibitem{bartol2004}
    Barr GD, Gaisser TK, Lipari P, Robbins S and Stanev T 2004 \textit{Phys. Rev. D} \textbf{70} 023006

\bibitem{honda2004}
    Honda M, Kajita T, Kasahara K and Midorikawa S 2004 \textit{Phys. Rev. D}
\textbf{70} 043008

\bibitem{corsika}
    Heck D, Knapp J, Capdevielle JN, Schatz G and Thouw T 1998
Tech. Rep. FZKA 6019 Forschungszentrum Karlsruhe

\bibitem{sdss}
    Stecker FW, Salamon MH, Done C and Sommers P 1992
\textit{Phys. Rev. Lett.} \textbf{66} 2697 (1991), \textbf{69} 2738(E)

\bibitem{sdss_revision}
    Stecker FW 2005 \textit{Phys.Rev. D} \textbf{72} 107301

\bibitem{mpr}
    Mannheim K, Protheroe RJ and Rachen JP 2000 \textit{Phys. Rev. D} \textbf{63}
023003

\bibitem{loeb_waxman_starburst}
    Loeb A and Waxman E 2006 \textit{J. Cosmol. Astropart. Phys.} JCAP05 003

\bibitem{zhv_charm}
    Zas E, Halzen F and V\'{a}zquez RA 1993 \textit{Astropart. Phys.} \textbf{1} 297

\bibitem{naumov_rqpm_a}
    Fiorentini G, Naumov A and Villante FL 2001 \textit{Phys. Lett. B} \textbf{510} 173

\bibitem{naumov_rqpm_b}
    Bugaev EV \emph{et al} 1989 \textit{Il Nuovo Cimento} 12C, No. 1, 41

\bibitem{martin_gbw}
    Martin AD, Ryskin MG and Stasto AM 2003 \textit{Acta Phys. Polon.} B34 3273

\end{thebibliography}

\section*{References}


\begin{thebibliography}{42}


\bibitem{wb}
Waxman~E and Bahcall~J 1997, \emph{Phys. Rev. Lett.} 78, 2292

\bibitem{crgrb2}
Wick~S, Dermer~C D and Atoyan~A 2004, \emph{Astropart. Phys.} 21, 125

\bibitem{batse}
Paciesas~W S et al. 1999, \emph{ApJS} 122, 465 (astro-ph/9903205)\\
http://www.batse.msfc.nasa.gov/batse/grb/catalog/

\bibitem{grbreview}
Mesz\'aros~P 2006, \emph{Rept.Prog.Phys.} 69, 2259

\bibitem{piranreview}
Piran~T 2005, \emph{Rev. Mod. Phys.} 76, 1143 

\bibitem{athar}
Athar~H, Kim~C S and Lee~J 2006, \emph{Mod. Phys. Lett.} A 21, 1049 

\bibitem{kashti}
Kashti~T and Waxman~E 2005, \emph{Phys. Rev. Lett.} 95, 181101

\bibitem{murase}
Murase~L and Nagataki~S 2006, \emph{Phys. Rev.} D 73, 063002

\bibitem{supranova} 
Razzaque~S, Mesz\'aros~P and Waxman~E 2003, \emph{Phys. Rev. Lett.} 90, 1103

\bibitem{precursor}
Mesz\'aros~P and Waxman~E 2001, \emph{Phys. Rev. Lett.} 87, 17110

\bibitem{afterglow}
Waxman~E and Bahcall~J 2000, \emph{ApJ} 541, 707

\bibitem{ama:nature-pub}
Andr\'{e}s~E et al. 2001, \emph{Nature} 410, 441

\bibitem{nu2004:ama}
Woschnagg~K et~al. 2005, \emph{Nuclear Physics} B \emph{Proc. Suppl.} 143, 343

\bibitem{ama:grb-muon}
Kuehn K et al. 2005, in Proc 29$^{\mathrm{th}}$ \emph{Int. Cosmic Ray Conf.} (in astro-ph/0509330)

\bibitem{ama:grb-casc}
Hughey B et al. 2005, in Proc 29$^{\mathrm{th}}$ \emph{Int. Cosmic Ray Conf.} (in astro-ph/0509330)

\bibitem{becker}
Becker J et al. 2006, \emph{Astropart.Phys.} 25, 118-128

\bibitem{guetta}
Guetta~D et al. 2004, \emph{Astropart. Phys.} 20, 429

\bibitem{icrc05:grb}
Stamatikos M et al. 2005, in Proc 29$^{\mathrm{th}}$ \emph{Int. Cosmic Ray Conf.} (in astro-ph/0509330)

\end{thebibliography}
 \end{document}